\documentclass[a4paper,11pt]{article}
\usepackage{jheppub}
\usepackage{mathtools}
\usepackage{physics}
\usepackage{orcidlink}
\usepackage{fontawesome5}
\graphicspath{{plots/}}

\newcommand{\gs}{g_\star}
\newcommand{\gss}{g_{\star s}}
\newcommand{\mr}{m_\varphi}
\newcommand{\Trh}{T_\text{rh}}
\newcommand{\arh}{a_\text{rh}}
\newcommand{\ard}{a_\phi}
\newcommand{\rgw}{\rho_\text{GW}}
\newcommand{\ogw}{\Omega_\text{GW}}
\newcommand{\rR}{\rho_\text{SM}}
\newcommand{\DNeff}{\Delta N_\text{eff}}
\newcommand{\GN}{G_\mathrm{N}}
\newcommand{\calM}{\mathcal{M}}
\newcommand{\Msqr}[1]{\abs{\calM_{#1}}^2}
\newcommand{\MPl}{M_P}

\title{Pre-thermalized Gravitational Waves}
\author[a]{Nicolás Bernal\,\orcidlink{0000-0003-1069-490X}}
\author[b]{, Quan-feng Wu\,\orcidlink{0000-0002-5716-5266}}
\author[b]{, Xun-Jie Xu\,\orcidlink{0000-0003-3181-1386}}
\author[c]{and Yong Xu\,\orcidlink{0000-0002-4582-8747}}
\affiliation[a]{New York University Abu Dhabi\\
	PO Box 129188, Saadiyat Island, Abu Dhabi, United Arab Emirates}
\affiliation[b]{Institute of High Energy Physics, Chinese Academy of Sciences, Beijing 100049, China}
\affiliation[c]{PRISMA$^+$ Cluster of Excellence and Mainz Institute for Theoretical Physics\\
	Johannes Gutenberg University, 55099 Mainz, Germany}
\emailAdd{nicolas.bernal@nyu.edu}
\emailAdd{wuquanfeng@ihep.ac.cn}
\emailAdd{xuxj@ihep.ac.cn}
\emailAdd{yonxu@uni-mainz.de}

\abstract{We investigate a novel gravitational wave (GW) production mechanism from gravitons generated during the pre-thermal phase of cosmic reheating, where the energy density is dominated by non-thermalized inflaton decay products, dubbed \emph{reheatons}. We consider multiple production channels, including: $i)$ pure inflaton-inflaton annihilation, $ii)$ graviton Bremsstrahlung from inflaton decay, $iii)$ scatterings between an inflaton and a reheaton, and $iv)$  scatterings among reheatons. To determine the resulting GW spectrum, we solve the Boltzmann equation to obtain the graviton phase-space distribution for each channel. We find that the third channel, $iii)$, dominates due to the large occupation number of reheatons at highly-energetic states during the pre-thermalization phase. Notably, in scenarios with a low inflaton mass, the GW spectrum could fall within the sensitivity range of future experiments such as the Einstein Telescope, the Cosmic Explorer, the Big Bang Observer, and ultimate DECIGO.}

\begin{document}
	\begin{flushright}
		MITP-25-022
	\end{flushright}
	\maketitle
	
	\section{Introduction}
	The early Universe hosts a variety of gravitational wave (GW) sources. In inflationary cosmology, the tensor perturbations generated during cosmic inflation provide one such source~\cite{Caprini:2018mtu}. Following inflation, the energy stored in the inflaton field must be transferred to other fields to reheat the Universe and establish a thermal radiation-dominated epoch with a temperature of at least $\mathcal{O}$(MeV) to ensure a successful Big Bang Nucleosynthesis (BBN)~\cite{Kawasaki:2000en, Hannestad:2004px, Cyburt:2015mya, deSalas:2015glj}. This energy transfer process, known as cosmic reheating~\cite{Allahverdi:2010xz, Amin:2014eta, Lozanov:2019jxc, Barman:2025lvk}, can itself be a source of GWs, depending on the specific setup of the reheating model and the relevant dynamics involved. In addition, GWs can be produced through non-perturbative preheating effects~\cite{Caprini:2018mtu}.
	
	In this work, we focus on GW production in the context of perturbative reheating. In this regard, several mechanisms have been explored in the literature, including $i)$ non-thermal graviton production processes such as graviton Bremsstrahlung~\cite{Nakayama:2018ptw, Huang:2019lgd, Barman:2023ymn, Barman:2023rpg, Kanemura:2023pnv, Bernal:2023wus, Tokareva:2023mrt, Hu:2024awd, Choi:2024acs, Barman:2024htg, Inui:2024wgj, Jiang:2024akb} and graviton pair production from inflaton annihilation~\cite{Ema:2015dka, Ema:2016hlw, Ema:2020ggo, Choi:2024ilx}, $ii)$ graviton production from inflaton scattering with thermalized daughter particles~\cite{Xu:2024fjl}, and $iii)$ graviton production from the scattering between thermalized daughter particles during reheating~\cite{Bernal:2024jim}. We also refer to Refs.~\cite{Ghiglieri:2015nfa, Ghiglieri:2020mhm, Ringwald:2020ist, Klose:2022knn, Ringwald:2022xif, Klose:2022rxh, Ghiglieri:2022rfp, Drewes:2023oxg, Ghiglieri:2024ghm} for studies of GW production from thermal plasma interactions in the radiation-dominated epoch after reheating. Previous studies such as Refs.~\cite{Xu:2024fjl, Bernal:2024jim} have assumed daughter fields that feature sizable gauge couplings with fast and instantaneous thermalization during reheating.
	
	In contrast to these earlier works, we investigate for the first time a novel GW production channel where gravitons are emitted during the {\it pre-thermal phase} of reheating. To this end, we assume the inflaton decay products to have small self-interactions, and to feebly communicate with the standard model (SM) particle content. Consequently, a slow thermalization is expected. This phase is particularly interesting because of the presence of numerous non-thermalized hard daughter particles with momenta of order of the inflaton mass. Since graviton production is an ultraviolet (UV) dominated process,\footnote{This is analogous to the UV freeze-in mechanism for dark matter production during pre-thermalization~\cite{Garcia:2018wtq, Chowdhury:2023jft}.} the production rate is expected to be higher than that from scatterings involving thermalized particles. Consequently, we anticipate an enhancement of the GW spectrum arising from this early phase, which we refer to as pre-thermalized GWs.\footnote{Note that here we specifically refer to GWs produced during the pre-thermalization phase. Moreover, the temperatures under consideration remain well below the Planck scale, ensuring that gravitons do not thermalize.} This constitutes the primary objective of the present study.
	
	Furthermore, we apply a systematic approach to study the GW spectrum by solving the Boltzmann equation at the phase-space distribution level, offering a robust cross-check with existing methods and results in the literature. This serves as another key objective of our work. Interestingly, the produced GW spectra turn out to be within the reach of next-generation GW observatories. In addition, since GWs propagate as dark radiation, they contribute to the effective number of neutrino species, $N_\text{eff}$. This contribution is also taken into account in the analysis.
	
	The remainder of this article is structured as follows. In Section~\ref{sec:setup}, we introduce the model setup. Section~\ref{sec:prethermalization} discusses the pre-thermal and thermal phases of reheating. In Section~\ref{sec:GWs}, we present the GW spectra, with particular emphasis on those produced during pre-thermalization, and provide a comparison with other sources. Finally, we summarize our main findings in Section~\ref{sec:conclusion}. Appendices~\ref{sec:M2} and~\ref{sec:collision} contain detailed computations of matrix elements and collision terms relevant to graviton production. The analytical derivation and the comparison with fully numerical results for the GW spectrum are presented in Appendix~\ref{sec:compare}.
	
	\section{The Model}\label{sec:setup}
	We consider a minimally coupled gravity framework, where the action is given by
	\begin{equation} \label{eq:action} 
		S \supset \int d^4x\, \sqrt{-g} \left[\mathcal{L}_{\text{EH}} + \mathcal{L}_\phi + \mathcal{L}_{\phi\varphi} + \mathcal{L}_{\varphi}\right].
	\end{equation}
	Here, $g$ denotes the determinant of the metric $g_{\mu\nu}$, $\mathcal{L}_{\text{EH}} = \frac12\, M_P^2\, R$ represents the Einstein-Hilbert term for gravity,\footnote{In this paper, we adopt the $(+, -, -, -)$ sign convention for the metric. In addition, we should clarify that the Riemann and Einstein signs in our setup are $-$ and $+$, respectively. Hence, there is no extra minus sign in the Einstein-Hilbert term $\mathcal{L}_{\text{EH}}$. For further clarification, see the $(-, -, +)$ convention in the ``table of sign conventions'' of Ref.~\cite{Misner:1973prb}. We have checked that the notation used in this work is the same as that in Refs.~\cite{Donoghue:1994dn, Choi:1994ax}.} and $R$ is the Ricci scalar, and $M_P \equiv 1/\sqrt{8 \pi\, \GN} \simeq 2.4 \times 10^{18}$~GeV is the reduced Planck mass. The second term in Eq.~\eqref{eq:action} corresponds to the Lagrangian density of a real singlet inflaton field $\phi$, given by
	\begin{equation} 
		\mathcal{L}_\phi = \frac12\, g^{\mu\nu}\, \partial_{\mu} \phi\, \partial_{\nu} \phi - V(\phi)\,, 
	\end{equation} 
	where $V(\phi)$ represents the inflaton potential. Although this potential may have a complicated form during inflation, it is assumed to be approximated by $V(\phi) \simeq \frac12\, m_\phi^2\, \phi^2$ after inflation, particularly during the oscillatory phase of the inflaton. This quadratic form occurs in various inflationary models, including Starobinsky inflation~\cite{Starobinsky:1980te}, certain classes of $\alpha$-attractor models~\cite{Kallosh:2013hoa, Kallosh:2013maa}, and both small- and large-field polynomial inflation scenarios~\cite{Drees:2021wgd, Bernal:2021qrl, Drees:2022aea, Bernal:2024ykj}. Depending on the specific inflationary model, the inflaton mass $m_\phi$ can vary significantly. For example, in the Starobinsky or large-field polynomial inflation model, it can be as large as $m_\phi \simeq 10^{13}$~GeV, whereas in a small-field polynomial inflationary scenario, it can be as low as $m_\phi \simeq 10^2$~GeV. In this work, we take an agnostic approach and treat the inflaton mass $m_\phi$ as a free parameter.
	
	The third term in Eq.~\eqref{eq:action} describes the interaction between the inflaton and a real bosonic daughter field $\varphi$, which eventually mediates the energy transfer from the inflaton to the SM degrees of freedom; $\varphi$ is therefore called the reheaton. We focus on inflaton decay via the interaction
	\begin{equation} \label{eq:phivarphi}
		\mathcal{L}_{\phi \varphi} \supset \frac12\, \mu\, \phi\, \varphi^2\,,
	\end{equation} 
	where $\mu$ is a dimension-one coupling constant. This trilinear term induces an inflaton decay rate into a pair of reheatons
	\begin{equation} \label{eq:Gammaphi}
		\Gamma_\phi = \frac{\mu^2}{32\pi\, m_\phi} \left[1 - \left(\frac{2\, \mr}{m_\phi}\right)^2\right]^{1/2} \simeq  \frac{\mu^2}{32\pi\, m_\phi} \,,
	\end{equation}
	where $\mr$ is the mass of $\varphi$.
	
	The final term in Eq.~\eqref{eq:action} describes the free-field Lagrangian density for $\varphi$, given by
	\begin{equation}
		\mathcal{L}_\varphi = \frac12\, g^{\mu\nu}\, \partial_{\mu} \varphi\, \partial_{\nu} \varphi - \frac12\, \mr^2\, \varphi^2\,.
	\end{equation} 
	Note that we have omitted any non-gravitational direct interactions between the inflaton and the SM degrees of freedom. Energy transfer to the SM sector occurs through the subsequent decays or annihilations of $\varphi$ into SM degrees of freedom, which will be discussed in Section~\ref{sub:to-SM}.
	
	To derive the gravitational interaction vertex and the corresponding Feynman rules for graviton production, we expand the metric $g_{\mu\nu}$ around the Minkowski metric $\eta_{\mu\nu} = (+,-,-,-)$ as~\cite{Donoghue:1994dn, Choi:1994ax}  
	\begin{equation}\label{eq:expansion}  
		g_{\mu\nu} = \eta_{\mu\nu} + \frac{2}{M_P} h_{\mu\nu} \,,  
	\end{equation}  
	where $h_{\mu\nu}$ represents the massless spin-2 graviton field, which has mass dimension one. From Eq.~\eqref{eq:expansion}, it follows that $\sqrt{-g} \simeq 1 + h/\MPl$, where $h\equiv h^{\mu}_{\mu}$ denotes the trace of the graviton field; it is zero in the transverse and traceless gauge (TT). Substituting this expansion into the action, we obtain the effective interaction between the graviton and the energy-momentum tensor~\cite{Donoghue:1994dn, Choi:1994ax}
	\begin{equation}\label{eq:effective}  
		\sqrt{-g}\, \mathcal{L} \supset -\frac{1}{M_P}\, h_{\mu\nu} \sum_k T_k^{\mu\nu}\,,  
	\end{equation}  
	where $T_k^{\mu\nu}$ denotes the energy-momentum tensor of a given particle species $k$, including the inflaton, the reheaton, and the degrees of freedom in the SM. In particular, graviton production requires a non-zero anisotropic energy-momentum tensor, meaning that the spatial components must satisfy $T^{ij} \neq 0$ for $i$, $j = 1$, 2, 3. The energy-momentum tensor of the inflaton takes the form  
	\begin{equation}\label{eq:Tmunu_inflaton}  
		T_\phi^{\mu\nu} = \partial^{\mu}\phi\, \partial^{\nu}\phi - \frac12\, \eta^{\mu\nu} \left(\partial_\alpha \phi\, \partial^{\alpha} \phi - m_\phi^2\, \phi^2\right),  
	\end{equation}  
	which implies that $T_\phi^{ij} = 0$ due to the homogeneity of the inflaton condensate, leading to vanishing anisotropic components. Consequently, the amplitude for any Feynman diagram featuring an outgoing graviton directly coupled to an {\it on-shell} inflaton is zero~\cite{Xu:2024fjl}.
	
	\section{Pre-thermalization and thermalization} \label{sec:prethermalization}
	\subsection{Inflatons after Inflation}
	Just after the end of the cosmic inflationary era at a scale factor $a = a_I$, the Universe is dominated by non-relativistic inflatons $\phi$ of mass $m_\phi$. Inflatons decay into pairs of ultra-relativistic reheatons $\varphi$ with a total decay width $\Gamma_\phi$ given in Eq.~\eqref{eq:Gammaphi}. The evolution of the inflaton number density $n_\phi$ is given by
	\begin{equation}
		\frac{dn_\phi}{dt} + 3\, H\, n_\phi = - \Gamma_\phi\, n_\phi\,,
	\end{equation}
	where $H$ is the Hubble expansion rate, which implies that
	\begin{equation} \label{eq:nphi0}
		n_\phi(t) = n_\phi(t_I) \left(\frac{a_I}{a(t)}\right)^3 e^{-\Gamma_\phi (t - t_I)},
	\end{equation}
	with $a$ the cosmic scale factor, $t_I$ and $a_I$ correspond to the time and scale factor at the end of the inflationary era, respectively. The value of $n_\phi(t_I)$ can be extracted from the scale of inflation $H_I$:
	\begin{equation}
		n_\phi(t_I) = \frac{3\, M_P^2\, H_I^2}{m_\phi}\,.
	\end{equation}
	It is interesting to note that, in a comoving frame, inflatons with momentum $p_\phi$ form a condensate characterized by a phase-space distribution
	\begin{equation} \label{eq:f-phi-delta}
		f_\phi(t, p_\phi) = (2 \pi)^3\, n_\phi(t)\, \delta^{(3)}(\vec p_\phi) = 2 \pi^2\, \frac{n_\phi(t)}{p_\phi^2}\, \delta(p_\phi)\,.
	\end{equation}
	
	Furthermore, taking into account that during this period the expansion of the Universe is dominated by non-relativistic inflatons,
	\begin{equation}
		H(a) = H_I \left(\frac{a_I}{a}\right)^\frac32,
	\end{equation}
	Eq.~\eqref{eq:nphi0} can be expressed as
	\begin{equation}
		n_\phi(a) = \frac{3\, M_P^2\, H_I^2}{m_\phi} \left(\frac{a_I}{a}\right)^3 \exp\left[- \gamma_{\phi} \left(\frac{a}{a_I}\right)^\frac32 \left[1 - \left(\frac{a_I}{a}\right)^\frac32\right]\right],
	\end{equation}
	where the dimensionless parameter $\gamma_{\phi}$ is defined as
	\begin{equation} \label{eq:gamma-phi}
		\gamma_{\phi} \equiv \frac23\, \frac{\Gamma_{\phi}}{H_I}\,.
	\end{equation}
	Given that for unitarity reasons $\mu \lesssim m_\phi$, it is expected that $\gamma_{\phi} \ll 1$. A small $\gamma_{\phi}$ also ensures that the inflaton decay becomes significant only after the end of the inflationary era and therefore has negligible effects on the slow-roll phase. Taking into account that in this era the Universe is dominated by non-relativistic inflatons, it follows that the evolution of the cosmic time is
	\begin{equation}
		t(a) = t_I + \frac{2}{3\, H_I}\left[\left(\frac{a}{a_I}\right)^{3/2} - 1\right].
	\end{equation}
	
	The period dominated by non-relativistic inflatons is followed by an era where ultra-relativistic reheatons (i.e. radiation domination) drive the Hubble expansion of the Universe. The transition occurs at a scale factor $a = \ard$. Although it is not an instantaneous event, one can define it at the time $t = t_I + \Gamma_{\phi}^{-1}$, corresponding to 
	\begin{equation} \label{eq:ardoaI}
		\frac{\ard}{a_I} = \left(\frac{1}{\gamma_{\phi}}+1\right)^{2/3} \simeq \gamma_\phi^{-\frac23}.
	\end{equation}
	From that moment on, ultra-relativistic reheatons start to dominate the energy density of the Universe. Their phase-space distribution is explored in the next section.
	
	\subsection{Non-thermal Reheatons}
	At production, reheatons are monochromatic with momentum $p_\varphi = m_\phi/2$. However, continuous production over time makes their distribution broad. The evolution of the phase-space distribution of the reheatons can be followed by the use of the Boltzmann transport equation
	\begin{equation} \label{eq:reheaton-dfdt}
		\frac{\partial f_\varphi}{\partial t} - H\, p_\varphi\, \frac{\partial f_\varphi}{\partial p_\varphi} = \Gamma_\varphi\,,
	\end{equation}
	where the collision term $\Gamma_\varphi$ is 
	\begin{align}
		\Gamma_\varphi(t, p_\varphi) &= \frac{1}{2 E_\varphi} \int d\Pi_{p_\phi}\, d\Pi_{p_{\varphi2}}\, (2\pi)^4\, \delta^{(4)}(p_\phi - p_\varphi - p_{\varphi2})\, |\mathcal{M}_{\phi \to \varphi \varphi}|^2\, f_\phi(t, p_\phi)\nonumber\\
		&= 16 \pi^2\, \frac{n_\phi\, \Gamma_\phi}{m_\phi^2}\, \delta\left(p_\varphi - \frac{m_\phi}{2}\right)
	\end{align}
	for ultra-relativistic reheatons, with $d\Pi_i$ being the Lorentz-invariant phase space, $E_\varphi$ the energy of the reheaton, and $|\mathcal{M}_{\phi \to \varphi \varphi}|^2 = \mu^2$ the squared amplitude for the decay $\phi \to \varphi\, \varphi$. Equation~\eqref{eq:reheaton-dfdt} can be analytically solved (see, e.g. Ref.~\cite{Wu:2024uxa}). The change of variable to the comoving momentum $\tilde p_\varphi \equiv a\, p_\varphi$ allows us to rewrite it as
	\begin{equation}
		\frac{df_\varphi(t, \tilde p_\varphi)}{dt} = \Gamma_\varphi\,,
	\end{equation}
	which implies that
	\begin{equation}
		f_\varphi(t, \tilde p_\varphi) = 32 \pi^2\, \frac{\Gamma_\phi}{H(\tilde a')}\, \frac{n_\phi(\tilde a')}{m_\phi^3}\, \tilde \Theta\left[\frac{m_\phi}{2}\, a_I \leq \tilde p_\varphi \leq \frac{m_\phi}{2}\, a(t)\right]
	\end{equation}
	for $\tilde a' \equiv 2\, \tilde p_\varphi / m_\phi$ and $\tilde\Theta$ the generalized Heaviside step function. In turn
	\begin{equation}
		f_\varphi(t, p_\varphi) = 32 \pi^2\, \frac{\Gamma_\phi}{H(a')}\, \frac{n_\phi(a')}{m_\phi^3}\, \tilde \Theta\left[\frac{m_\phi}{2}\, \frac{a_I}{a(t)} \leq p_\varphi \leq \frac{m_\phi}{2}\right]
	\end{equation}
	for $a' \equiv 2\, p_\varphi\, a(t) / m_\phi$ which, in terms of the scale factor, becomes 
	\begin{align} \label{eq:f-reheaton}
		f_\varphi(a, p_\varphi) &= 96 \pi^2\, \frac{M_P^2\, H_I\, \Gamma_\phi}{m_\phi^4} \left(\frac{m_\phi}{2 p_\varphi} \frac{a_I}{a}\right)^\frac32 \exp\left[\gamma_\phi \left(1 - \left(\frac{2 p_\varphi}{m_\phi}\frac{a}{a_I}\right)^\frac32\right)\right]\nonumber\\
		&\qquad\qquad \times \tilde \Theta\left[\frac{m_\phi}{2}\, \frac{a_I}{a} \leq p_\varphi \leq \frac{m_\phi}{2}\right].
	\end{align}
	
	The reheaton number density can be expressed in the exact form,
	\begin{equation} \label{eq:-1}
		n_\varphi(t) = \frac{1}{2 \pi^2} \int f_\varphi(a, p_\varphi)\, p_\varphi^2\, dp_\varphi = 6\, \frac{M_P^2\, H_I^2}{m_\phi} \left(\frac{a_{I}}{a(t)}\right)^3 \left[1 -e^{-\Gamma_\phi\, (t-t_I)}\right],
	\end{equation}
	which can also be obtained from particle number conservation (that is, $(n_\phi + n_\varphi/2)\, a^3$ should be constant). The latter serves as a useful cross-check of Eq.~\eqref{eq:f-reheaton}.
	
	Let us comment here that it could be possible for $\varphi$ to reach kinetic equilibrium or even chemical equilibrium, dramatically changing the phase-space distribution previously computed. However, the former requires a sufficiently high elastic scattering rate of reheatons $\varphi \varphi \to \varphi \varphi$, while the latter requires a particle-number-changing process such as $\varphi \varphi \to \varphi \varphi \varphi \varphi$. We have estimated the rates and find that they are in general below the Hubble expansion rate for the trilinear coupling strengths considered in this work.
	
	\subsection{Standard Model Thermal Bath} \label{sub:to-SM}
	After inflatons decay to ultra-relativistic reheatons, the Universe becomes radiation dominated. At a later epoch, they eventually release all their energy into SM particles through decays or annihilations, creating the SM thermal bath. As reheatons are much lighter than inflatons, they are expected to decay at a much later time.\footnote{Note that the decay or annihilation rate compared to the Hubble expansion rate is typically suppressed when the Universe is at a temperature much higher than all mass scales involved in the decay or annihilation process. For example, if $\varphi$ has a small (compared to $H_I$ and $m_\phi$) mass $\mr$ above the electroweak scale, its decay rate is typically below ${y^2\, \mr}/{(8\pi)}$, with $y$ being a dimensionless Yukawa coupling to SM fermions.} Let us define $a = \arh$ the scale factor at the equality between the energy density of reheatons and the SM: $\rho_\varphi(\arh) = \rho_{\rm SM}(\arh)$, where
	\begin{equation}
		\rho_{\rm SM}(T) = \frac{\pi^2}{30}\, \gs(T)\, T^4,
	\end{equation}
	with $\gs$ corresponding to the number of relativistic degrees of freedom contributing to $\rho_{\rm SM}$. Therefore, the reheating temperature $\Trh$ can be defined as $\Trh \equiv T(\arh)$. As the present analysis focuses on GWs with a non-thermal origin, the details of the SM thermal era are irrelevant, as long as the transition from $\varphi$ to the SM content occurs when $\varphi$ is {\it relativistic}.
	
	The duration of the era dominated by reheatons can be computed by the use of the scaling of the Hubble rate $H(\ard)\, \ard^2 = H(\arh)\, \arh^2$, which translates into
	\begin{equation} \label{eq:arhoard}
		\frac{\arh}{\ard} = \left[\frac{3}{\pi}\, \sqrt{\frac{10}{\gs(\Trh)}}\, \frac{M_P\, H_I}{\Trh^2} \left(\frac{a_I}{\ard}\right)^\frac32\right]^\frac12.
	\end{equation}
	Furthermore, after the Universe fully transitions to the SM thermal plasma, the subsequent evolution follows the standard cosmological model in which the scale factor can be determined using the SM entropy conservation. In particular, between $T = \Trh$ and the present CMB temperature $T_0 \simeq 2.73~\text{K} \simeq 0.23 \times 10^{-12}$~GeV one has that
	\begin{equation} \label{eq:a0oarh}
		\frac{a_0}{\arh} = \left[\frac{\gss(\Trh)}{\gss(T_0)}\right]^{1/3} \frac{\Trh}{T_0}\,,
	\end{equation}
	where $a_0$ is the scale factor at present, and $\gss(T)$ accounts for the effective degrees contributing to the SM entropy. Combining Eqs.~\eqref{eq:ardoaI}, \eqref{eq:arhoard} and~\eqref{eq:a0oarh}, one find that
	\begin{align} \label{eq:aI-a0-ratio}
		\frac{a_0}{a_I} &= \left(\frac{3}{\pi}\right)^\frac12 \left(\frac{10}{\gs(\Trh)}\right)^\frac14 \left(\frac{\gss(\Trh)}{\gss(T_0)}\right)^\frac13 \left(\frac{1 + \gamma_\phi}{\gamma_\phi}\right)^\frac16 \frac{\sqrt{M_P\, H_I}}{T_0} \nonumber\\
		&\simeq 1.6 \times \left(\frac{1 + \gamma_\phi}{\gamma_\phi}\right)^\frac16 \frac{\sqrt{M_P\, H_I}}{T_0}
	\end{align}
	which, as expected, is independent of $\Trh$ (as long as $\Trh$ is higher than the electroweak scale).
	
	Given that $\arh \geq \ard$ within our model setup, we have the upper bound on $\Trh$
	\begin{align} \label{eq:Trh-below}
		\Trh  & \lesssim \sqrt { \frac {2} {\pi} } \left( \frac {10} {\gs (\Trh)} \right)^{1/4} \sqrt {M_P\, \Gamma_\phi}\nonumber\\
		&\simeq 2.2 \times 10^{13} \left(\frac{10^{13}~\text{GeV}}{m_\phi}\right)^{1/2} \left(\frac{\mu}{10^{12}~\text{GeV}}\right) \text{GeV}.
	\end{align}
	The current upper bound on the reheating temperature is $\Trh \lesssim 5.5 \times 10^{15}~\text{GeV}$, which follows from the constraint on the inflationary Hubble scale, $H_I \lesssim 2.0 \times 10^{-5}~M_P$, as inferred from the BICEP/Keck 2018 results~\cite{BICEP:2021xfz}. We note that in the limit where $\arh \to \ard$, that is, in the instantaneous decay scenario, we reproduce the usual result of $\Trh$. A lower bound on $\Trh \geq 4$~MeV also exists, coming from BBN~\cite{Kawasaki:2000en, Hannestad:2004px, deSalas:2015glj, Cyburt:2015mya, Hasegawa:2019jsa}.
	
\begin{figure}
		\centering
		\includegraphics[width=0.6\textwidth]{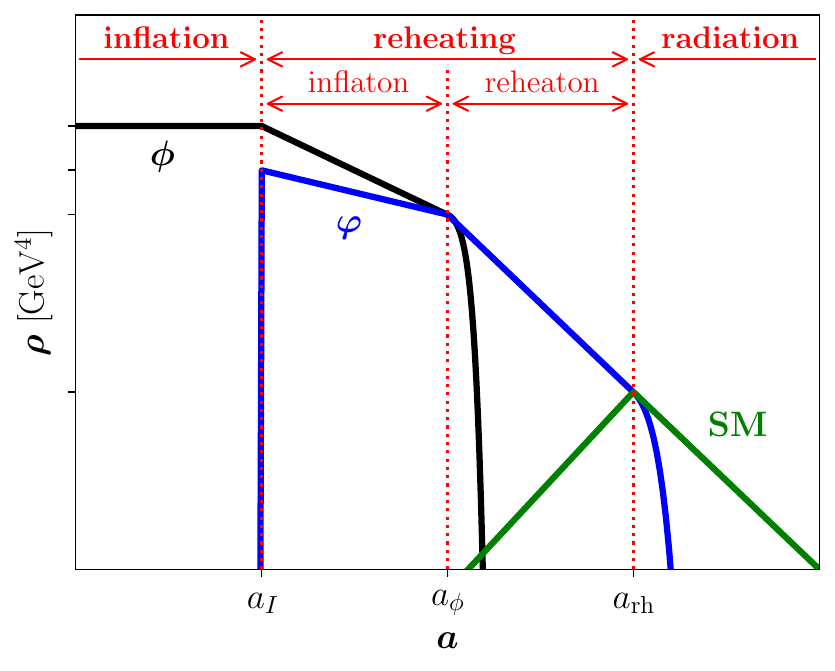}
		\caption{Artistic view of the evolution of the energy density of the different components of the Universe: inflaton $\phi$ in black, reheaton $\varphi$ in blue, and SM in green. Cosmic inflation corresponds to $ a < a_I$. Cosmic reheating occurs when $a_I < a < \arh$, first dominated by the inflaton ($a_I < a < \ard$) and then by the reheaton ($\arh < a < \arh$). Finally, the early Universe is dominated by SM radiation at $\arh < a$.}
		\label{fig:history}
	\end{figure}

	A sketch of the evolution of the energy density of the different components of the Universe is shown in Fig.~\ref{fig:history}: inflatons $\phi$ are shown in black, reheatons $\varphi$ in blue, and SM in green. Cosmic inflation corresponds to $ a < a_I$. Cosmic reheating occurs when $a_I < a < \arh$, first dominated by the inflaton ($a_I < a < \ard$) and then by the reheaton ($\arh < a < \arh$). Finally, the Universe is dominated by SM radiation at $\arh < a$, until the matter-radiation equality at $T \simeq 0.8$~eV.
	
	\section{Gravitational Waves} \label{sec:GWs}
\begin{figure}
		\centering
		\includegraphics[width=1\textwidth]{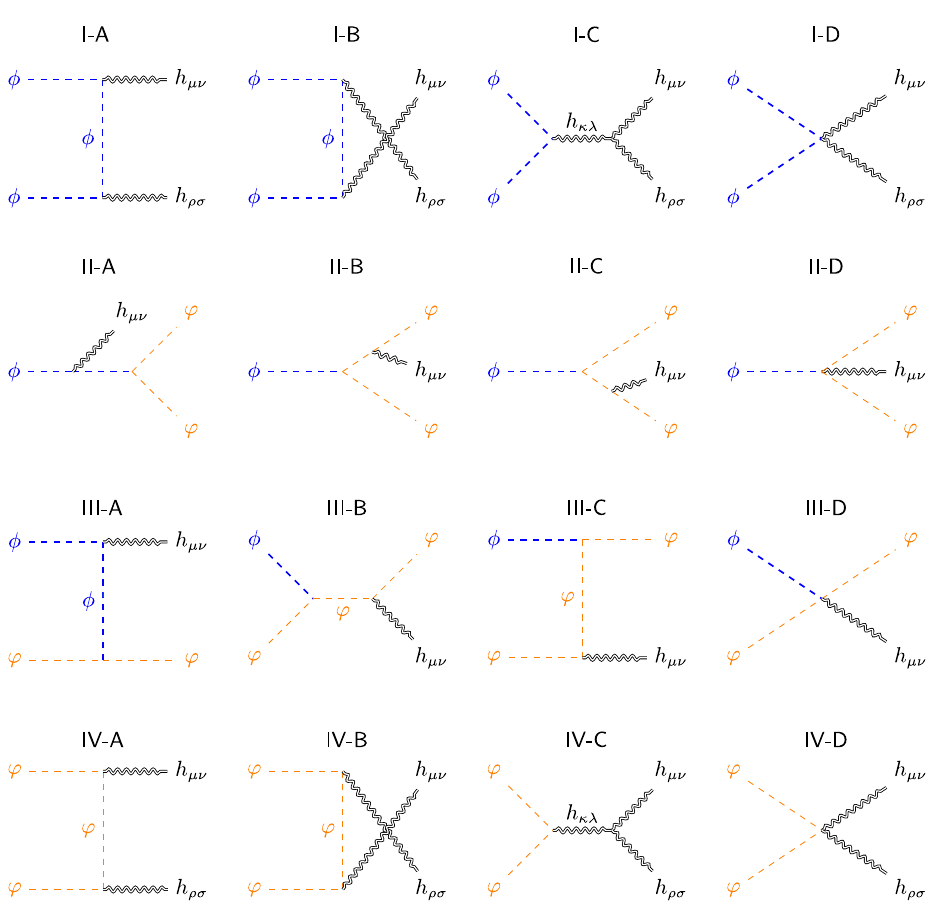}
		\caption{Feynman diagrams responsible for graviton ($h_{\mu\nu}$) production from inflatons ($\phi$) and reheatons ($\varphi$), in the presence of the trilinear interaction in Eq.~\eqref{eq:phivarphi}.}
		\label{fig:Feyn-all} 
	\end{figure}

	In this work, we examine the graviton production from both inflatons and reheatons. The corresponding Feynman diagrams are presented in Fig.~\ref{fig:Feyn-all}. The first row (I) depicts graviton-pair production from inflaton annihilation. The second row (II) illustrates single-graviton production via graviton Bremsstrahlung during inflaton decay into a pair of reheatons. The third row (III) represents inflaton-reheaton scattering, leading to the emission of a single graviton. Finally, the last row (IV) shows the production of graviton pairs from reheaton annihilations.\footnote{We remind the reader that in the current setup with a minimal coupled Einstein-Hilbert action, there is no vertex between a single particle and two gravitons.}  This corresponds to all possible tree-level diagrams with at most four external particles. 
	
	We note that, beyond tree level, graviton production can also arise at loop level. For example, one can have the semi-annihilation of inflatons into an inflaton and a graviton, or the inflaton decay into a pair of gravitons. Processes with more external particles can also occur, such as 2-to-3 annihilations $\phi\, \phi \to \phi\, \phi\, h$, $\varphi\, \varphi \to \varphi\, \varphi\, h$, or $\phi\, \varphi \to \phi\, \varphi\, h$. However, these production channels are further suppressed compared to direct inflaton annihilation~\cite{Xu:2024fjl}. 
	
	The production of GWs from all the processes can be computed by solving the Boltzmann equation for the full phase-space distribution of gravitons
	\begin{equation} \label{eq:f-boltz}
		\frac{\partial f_h}{\partial t} - H\, p_h\, \frac{\partial f_h}{\partial p_h} = \Gamma_h\,,
	\end{equation}
	where $f_h = f_h(a, p_h)$ is the phase-space distribution of the graviton $h_{\mu\nu}$ and $\Gamma_h$ denotes the production rate of $h_{\mu\nu}$. The backreaction terms that account for the absorption of $h_{\mu\nu}$ via the inverse processes of those presented in Fig.~\ref{fig:Feyn-all} are not included, because they are negligible as long as $f_h \ll 1$. Consequently, the right-hand side of Eq.~\eqref{eq:f-boltz} is independent of $f_h$, allowing us to rewrite Eq.~\eqref{eq:f-boltz} into the integral form as
	\begin{equation} \label{eq:f-from-int}
		f_h(a, p_h) = \int_{a_I}^a \frac{da'}{a'\, H(a')}\, \Gamma_h\left(a', p_h\, \frac{a}{a'}\right).
	\end{equation}
	It follows that the total energy density carried by GWs is
	\begin{equation} \label{eq:rhoGW-int}
		\rgw(a) = g_h \int  \frac{4\pi\, p_h^3\, dp_h}{(2\pi)^3}\, f_h(a, p_h)\,,
	\end{equation}
	where $g_h =2$ denotes the two degrees of freedom for massless gravitons. The GW spectrum at present (that is, at $a = a_0$) is
	\begin{equation}
		\ogw(f) \equiv \frac{1}{\rho_c}\, \frac{d\rgw(a_0)}{d\ln p_h} = 16 \pi^2\, \frac{f^4}{\rho_c}\, f_h(a_0, 2\pi\, f)\,,
	\end{equation}
	as a function of the frequency $f$ of the GW, with $\rho_c \simeq 1.05 \times 10^{-5}~h^2$~GeV/cm$^3$ the critical energy density at present~\cite{ParticleDataGroup:2024cfk}.
	
	For a generic process, $X_{1}+X_{2}+\cdots+X_{n}\to X_{n+1}+X_{n+2}+\cdots+X_{n+m}+h$,
	the production rate (also known as the collision term) involves the $(n+m)$-body phase space integral
	\begin{equation} \label{eq:collision-def}
		\Gamma_h \equiv \frac{1}{2 E_h}\int d\Pi_{1} \cdots d\Pi_{n+m}\, f_{1} \cdots f_{n}\, \Msqr{}\, (2\pi)^4\, \delta^{(4)}\left(\sum_{i=1}^{n} p_i - \sum_{j=n+1}^{n+m} p_j\right),
	\end{equation}
	where $d\Pi_i\equiv \frac{d^{3}p_i}{(2\pi)^{3}2E_i}$ is the Lorentz-invariant phase space with $p_i$ and $E_i$ the momentum and energy of the $i^\text{th}$ particle in the process, $f_i$ is the phase-space distribution of the $i^\text{th}$ particle, and $(2\pi)^4 \delta^{(4)}(\dots)$ standards for the common delta function responsible for energy-momentum conservation. 
	
	Given the above formulae, it is possible to compute $f_h$ and $\rgw$ for the processes presented in Fig.~\ref{fig:Feyn-all}. In the following, we discuss the production of GW in these processes and summarize the corresponding matrix elements and the production rates in Table~\ref{tab:M2andGamma}.
	\begin{table*}
		\centering
		\begin{tabular}{ccccc}
			\hline 
			Processes & Diagrams & $S$ & $\Msqr{}$ & $\Gamma_h$\\[2mm]
			\hline
			$\phi\phi\to hh$ & I & $\frac{1}{4}$ & $\frac{2\, m_{\phi}^{4}}{M_P^{4}}$ & $\frac{\pi\, \Msqr{}\, n_{\phi}^2}{16\, E_h^2\, m_{\phi}^2}\, \delta(E_h-m_{\phi})$ \\[2mm]
			$\phi\to\varphi\varphi h$ & II & $\frac{1}{2} $ & $\frac{2\, \mu^2}{M_P^2} \left(1 - \frac{m_{\phi}}{2\, E_{h}}\right)^2$ & $\frac{\Msqr{}\, n_{\phi}}{64\pi\, m_{\phi}\, E_h}$  \\[2mm]
			$\phi\varphi\to h\varphi$ & III & $1$ & $\frac{2\,\mu^2}{M_P^2} \left(1 - \frac{m_{\phi}}{2\, E_{h}}\right)^2$ & Eq.~\eqref{eq:x-8}\\[2mm]
			$\varphi\varphi\to hh$ & IV & $\frac{1}{4} $ & $\frac{2\, t^2\, (s+t)^2}{M_P^{4}\,s^2}$ & $\frac{\pi\, n_{\varphi}^{2}}{320\, E_{h}\, M_{P}^{4}}$  \\[2mm]
			\hline 
		\end{tabular}
		\caption{Squared amplitudes $\Msqr{}$, symmetry factors $S$ and production rates $\Gamma_h$ of gravitons for the processes presented in Fig.~\ref{fig:Feyn-all}.
			\label{tab:M2andGamma}}
	\end{table*}
	
	\subsection[\texorpdfstring{$\phi\, \phi \to h\, h$}{ϕ ϕ → h h}]{\boldmath $\phi\, \phi \to h\, h$}
	At the end of inflation, the condensate of non-relativistic inflatons can directly produce gravitons via diagrams I-A to I-D in Fig.~\ref{fig:Feyn-all}. The squared amplitude of this process can be obtained by taking the non-relativistic limit of the general calculation of GW production from scalar annihilation\footnote{We note that the process $\phi \phi \to \phi h$ is less suppressed than $\phi \phi \to h h$ in terms of coupling strength. However, for {\it non-relativistic} inflatons, the matrix element for $\phi \phi \to \phi h$ vanishes due to conservation of the angular momentum.} (see Eq.~\eqref{eq:M2-anni-nr}) and is~\cite{Choi:2024ilx}
	\begin{equation} \label{eq:M2-anni-nrB}
		\Msqr{\phi\phi\to hh} = \frac{2\, m_\phi^4}{M_P^4}\,;
	\end{equation}
	the symmetry factor of the final and initial state is $S = 1/4$.
	
	Since the squared amplitude is a constant, it can be moved outside the phase space integrals in Eq.~\eqref{eq:collision-def}. Labeling the particles in $\phi\phi\to hh$  from the left to the right by numbers 1 to 4, we compute $\Gamma_h$ as follows:
	\begin{align} \label{eq:Gamma-I}
		\Gamma_h &=\frac{S}{2 E_h}\int d\Pi_{1}\, d\Pi_{2}\, d\Pi_{3}\, f_\phi(p_1)\, f_\phi(p_2)\, 2\, \Msqr{\phi\phi\to hh}\, (2\pi)^4\, \delta^{(4)}(p_1 + p_2 - p_3 - p_4)\nonumber \\
		& =\frac{2\, S\, \Msqr{\phi\phi\to hh}}{2 E_h}\int d\Pi_{1}d\Pi_{2}\, \frac{1}{2 (E_1 + E_2 - E_h)}\, f_\phi(p_1)\, f_\phi(p_2)\, 2\pi\, \delta(E_1 + E_2 - 2 E_h)\nonumber \\
		& =\frac{\pi\, S\, \Msqr{\phi\phi\to hh}\, n_{\phi}^2}{8 E_h^2\, m_{\phi}^2}\, \delta(E_h - m_{\phi}) = \frac{\pi}{16}\, \frac{n_\phi^2}{M_P^4}\, \delta(E_h - m_{\phi})\,.
	\end{align}
	Here $\Msqr{\phi\phi\to hh}$ in the first step is multiplied by a factor of two to account for the double graviton production; and in the last step we have used the cold condensate distribution in Eq.~\eqref{eq:f-phi-delta} for $f_\phi$.
	
	The corresponding phase-space distribution for the produced gravitons at the end of the inflaton-dominated era is
	\begin{align} \label{eq:fh-infinf}
		f_h(\ard, p_h) &= \int_{a_I}^{\ard} \frac{da'}{a'\, H(a')}\, \Gamma_h\left(a', p_h\, \frac{\ard}{a'}\right) \nonumber\\
		&= \frac{9 \pi}{16}\, \frac{H_I^3}{m_\phi^2} \int_{a_I}^{\ard} \frac{da'}{a'} \delta\left(p_h\, \frac{\ard}{a'} -  m_\phi\right) \left(\frac{a_I}{a'}\right)^\frac92 \exp\left[- \frac43\, \frac{\Gamma_\phi}{H_I} \left(\frac{a'}{a_I}\right)^\frac32 \left[1 - \left(\frac{a_I}{a'}\right)^\frac32\right]\right] \nonumber\\
		&= \frac{9 \pi}{16} \left(\frac{H_I}{m_\phi}\right)^3 \left(\frac{a_I}{\ard}\, \frac{m_\phi}{p_h}\right)^\frac92 \exp\left[- \frac43\, \frac{\Gamma_\phi}{H_I} \left[\left(\frac{\ard}{a_I}\, \frac{p_h}{m_\phi}\right)^\frac32 - 1\right]\right] \nonumber\\
		&\qquad\qquad \times \tilde \Theta\left(m_\phi\, \frac{a_I}{\ard} \leq p_h \leq m_\phi\right),
	\end{align}
	which allows to compute the spectrum of GWs. Furthermore, the total energy density in GWs at $a = \ard$
	\begin{align}
		\rgw(\ard) &= \int \frac{p_h^3\, dp_h}{2\pi^2}\, f_h(\ard, p_h) \simeq \frac{9}{32 \pi}\, H_I^3 \left(\frac{a_I}{\ard}\right)^3 \int_{m_\phi \frac{a_I}{\ard}}^{m_\phi} dp_h \left(\frac{a_I}{\ard}\, \frac{m_\phi}{p_h}\right)^\frac32 \nonumber\\
		&\simeq \frac{9}{16 \pi}\, H_I^3\, m_\phi \left(\frac{a_I}{\ard}\right)^4,
	\end{align}
	in agreement with Eq.~(21) in Ref.~\cite{Choi:2024ilx}, and Eq.~(5.14) in Ref.~\cite{Xu:2024fjl} where a different method has been applied. The corresponding GW spectrum at present is
	\begin{equation} \label{eq:ogw_inflaton_inflaton}
		\ogw h^2(f) \simeq 5.66 \times 10^{-20} \left( \frac{\mu}{10^{12}~\text{GeV}}\right)^{3/2}  \left( \frac{m_\phi}{10^{13}~\text{GeV}}\right)^{3/4}  \left( \frac{f}{10^{7}~\text{Hz}}\right)^{-1/2}.
	\end{equation}
	The graviton energy at production is the same as the inflaton mass, namely $E_h =m_\phi$. Once redshift is taken into account, the GW spectrum spans over the frequency band
	\begin{equation} \label{eq:frequency_band}
		f_\text{min} \lesssim  f \lesssim f_\text{max} \,,
	\end{equation}
	where
	\begin{align}\label{eq:fre}
		f_\text{min} &\simeq 6.4 \times 10^6 \left(\frac{m_\phi}{10^{13}~\text{GeV}}\right)^{5/6} \left(\frac{\mu}{10^{12}~\text{GeV}}\right)^{1/3}\left(\frac{2.0 \times 10^{-5}\,M_P}{H_I}\right)^{2/3} \text{Hz} \\
		f_\text{max} &\simeq 1.1 \times 10^{10} \left(\frac{m_\phi}{10^{13}~\text{GeV}}\right)^{3/2} \left(\frac{10^{12}~\text{GeV}}{\mu}\right) \text{Hz}.
	\end{align}
	The GW spectrum with frequency above $f_\text{max}$ features a suppression factor $\sim e^{-(f/f_\text{max})^2}$ due to the exponential suppression of the inflaton number density. We note that in a scenario where the inflaton decays into the reheaton directly reheats the Universe, the coupling can be expressed as a function of $\Trh$. In such a case, Eq.~\eqref{eq:ogw_inflaton_inflaton} reproduces the results given in Eq.~(25) of Ref.~\cite{Choi:2024ilx} and Eq.~(6.9) of Ref.~\cite{Xu:2024fjl}. Finally, we note that this is a purely gravitational channel and, therefore, its matrix element and corresponding collision term are independent of the trilinear coupling $\mu$ (cf. Eqs.~\eqref{eq:M2-anni-nrB} and~\eqref{eq:Gamma-I}). However, its induced GW spectrum does have a dependence on $\mu$ through the redshift of the produced gravitons; cf. Eq.~\eqref{eq:ogw_inflaton_inflaton}.
	
	\subsection[\texorpdfstring{$\phi \to \varphi\,\varphi\, h$}{ϕ → φ φ h}]{\boldmath $\phi \to \varphi\,\varphi\, h$}
	When an inflaton $\phi$ decays into reheatons $\varphi$, gravitons are unavoidably produced via Bremsstrahlung, as shown in II-A to D in Fig.~\ref{fig:Feyn-all}. These diagrams lead to the following squared matrix element (see Eq.~\eqref{eq:M2-decay-final})
	\begin{equation} \label{eq:M2-phi-decay}
		\Msqr{\phi \to \varphi\varphi h} = \frac{2\, \mu^2}{M_P^2} \left(1 - \frac{m_\phi}{2\, E_h}\right)^2, 
	\end{equation}
	with the graviton energy $E_h$ in the range $0 < E_h < \frac{m_\phi}{2}$; for higher energies the process is kinematically forbidden. 
	
	Since $\Msqr{\phi \to \varphi\varphi h}$ in this process depends only on $E_h$, it can be moved outside the phase-space integral, similar to the calculation of $\Gamma_{h}$ in the previous subsection. Labeling the particles in $\phi\to\varphi\, \varphi\, h$ from the left to the right by numbers 1 to 4, $\Gamma_{h}$ is then given by 
	\begin{equation}
		\Gamma_{h} = \frac{\Msqr{\phi \to \varphi\varphi h}}{64 \pi}\, \frac{n_\phi}{m_\phi\, p_h}\,,
	\end{equation}
	see Section~\ref{sec:Gh2} for further details.
	
	The phase-space distribution for the produced gravitons at the end of the inflaton-dominated era is
	\begin{equation} \label{eq:fh-Brems}
		f_h(\ard, p_h) \simeq \frac{1}{64 \pi}\, \frac{\mu^2\, H_I}{p_h^3} \left(\frac{a_I}{\ard}\right)^\frac32,
	\end{equation}
	while the total energy density stored in graviton at $a = \ard$ is
	\begin{equation}
		\rgw(\ard) = \int_0^\frac{m_\phi}{2} \frac{p_h^3\, dp_h}{\pi^2}\, f_h(\ard, p_h) \simeq \frac{1}{128 \pi^3}\, m_\phi\, \mu^2\, H_I \left(\frac{a_I}{\ard}\right)^\frac32,
	\end{equation}
	which reproduces the result in Eq.~(3.13) of Ref.~\cite{Barman:2023ymn} in the limit $\ard \to \arh$, or in a scenario where inflatons directly decay into SM particles. The analytical expression for the GW can be written as
	\begin{equation} \label{eq:GWBrem2}
		\ogw h^2(f) \simeq 9.77\times 10^{-21}    \left(\frac{\mu}{10^{12}~\text{GeV}} \right)  \left(\frac{m_\phi}{10^{13}~\text{GeV}} \right)^\frac12 \left(\frac{f}{10^{7}~\text{Hz}} \right),
	\end{equation}
	with $f \lesssim f_\text{peak} \simeq f_\text{max}/2$. It is interesting to note that, at its peak, the amplitude of the spectrum $\ogw(f_\text{peak}) \propto m_\phi^2$, and only depends on $m_\phi$. This implies that the lower $m_\phi$, the smaller the GW amplitude would be. Similarly as in the previous case, higher frequency gravitons with $f > f_\text{peak}$ can also be generated, but the corresponding amplitude of the spectrum is exponentially suppressed. Finally, we note that in the case where reheating occurs through the coupling $\mu\, \phi\, \varphi^2$ (that is, $\varphi$ is identified with the SM Higgs), Eq.~\eqref{eq:GWBrem2} reproduces the result presented in Eq.~(4.5) of Ref.~\cite{Barman:2023ymn}. 
	
	\subsection[\texorpdfstring{$\phi\, \varphi \to \varphi\, h$}{ϕ φ → φ h}]{\boldmath $\phi\, \varphi \to \varphi\, h$}
	Gravitons can also be produced in 2-to-2 scatterings of inflatons with reheatons, in a reheaton-catalyzed inflaton-to-graviton conversion. The corresponding diagrams are depicted in the third row of Fig.~\ref{fig:Feyn-all}. The detailed computation of the squared amplitude is presented in Appendix~\ref{sec:phi_varphi}, where we explicitly show that the squared amplitude of $\phi\varphi\to \varphi h$ is identical to that of $\phi\to \varphi \varphi h$, as expected from crossing symmetries.
	
	The calculation of $\Gamma_{h}$ for $\phi\varphi\to\varphi h$ is similar to the calculation for $\phi\to\varphi\varphi h$ in the previous subsection; however, in this case one has 
	\begin{equation}\label{eq:Gamma_phivarphi}
		\Gamma_ \sim \frac{\pi}{4}\, \frac{\Msqr{}\, n_\phi\, n_\varphi}{E_h^2\, m_\phi^3}\, \tilde\Theta\left[\frac12 < \frac{E_h}{m_\phi} < 1\right],
	\end{equation}
	see Appendix~\ref{sec:Gh3} for further details. The corresponding phase-space distribution for the produced gravitons at the end of the inflaton-dominated era can be approximated as
	\begin{equation} \label{eq:fh_phivarphi}
		f_h(\ard, p_h) \simeq   \frac{\pi}{4}\, \frac{M_P^2\, \mu^2\, H_I^2\, \Gamma_\phi}{m_\phi^4\, p_h^3} \left(\frac{a_I}{\ard}\right)^3 
	\end{equation}
	for $\frac{m_\phi}{2}\, \frac{a_I}{\ard} <p_h < m_\phi$, which corresponds to a total energy density in the form of gravitons at $a = \ard$
	\begin{equation}
		\rgw(\ard) \simeq \frac{1}{4\pi} \frac{M_P^2\, \mu^2\, H_I^2\, \Gamma_\phi}{m_\phi^3} \left(\frac{a_I}{\ard}\right)^3.
	\end{equation}
	
	The GW spectrum is estimated as
	\begin{equation}\label{eq:GWphivarphi}
		\ogw h^2(f) \simeq 5.86 \times 10^{-16} \left( \frac{\mu}{10^{12}~\text{GeV}}\right)^5  \left( \frac{10^{13}~\text{GeV}}{m_\phi}\right)^\frac{11}{2}  \left( \frac{f}{10^7~\text{Hz}}\right),
	\end{equation}
	which also features a linear dependence on $f$ in the low frequency regime, similar to the Bremsstrahlung spectrum. At peak, the spectrum $\ogw(f_\text{peak}) \propto (\mu/m_\phi)^4$, being constant if $\mu \propto m_\phi$. Similarly to the previous case, the frequency is in the range $2\, f_\text{min} \lesssim f \lesssim f_\text{max}$.
	
	Before closing this section, we emphasize that the GW spectrum previously computed corresponds to a {\it non-thermal} reheaton. For a thermalized reheaton, the spectrum starts at lower frequencies and has a much suppressed amplitude, due to the change in the phase-space distribution~\cite{Xu:2024fjl}.
	
	\subsection[\texorpdfstring{$\varphi\, \varphi \to h\, h$}{φ φ → h h}]{\boldmath $\varphi\, \varphi \to h\, h$}
	Pairs of gravitons can also be produced by annihilations of ultra-relativistic reheatons, see the fourth line of Fig.~\ref{fig:Feyn-all}. In the ultra-relativistic limit, the squared amplitude in Eq.~\eqref{eq:M2-anni} reduces to~\cite{Ghiglieri:2022rfp}
	\begin{equation}
		\Msqr{\varphi\varphi\to hh} = \frac{2\, t^2\, (s + t)^2}{M_P^4\, s^2}\,.
	\end{equation}
	An exact analytical expression for the production rate $\Gamma_h$ for this channel is challenging to obtain due to the integration over the phase space. However, we estimate it to be
	\begin{equation} \label{eq:x-9-1}
		\Gamma_{h} \simeq \frac{\pi}{320}\, \frac{n_{\varphi}^{2}}{M_{P}^{4}\, E_{h}}\,,
	\end{equation}
	where $\frac{m_\phi}{2}\, \frac{a_I}{a} < E_h < \frac{m_\phi}{2} \times \min(1, a_\phi/a)$. The exact collision term $\Gamma_h$ can always be obtained numerically. The phase-space distribution of gravitons at $a =\ard$ is
	\begin{equation} \label{eq:fh_varphivarphi}
		f_h(\ard, p_h) \simeq \frac{9 \pi}{140}\, \frac{H_I\, \Gamma_\phi^2}{m_\phi^2\, p_h}\, \frac{a_I}{\ard},
	\end{equation}
	for $\frac{m_\phi}{2}\, \frac{a_I}{a} < p_h < \frac{m_\phi}{2} \times \min(1, 
	\ard/a)$. The total energy density for GW can be estimated as
	\begin{equation}
		\rgw(\ard) \simeq \frac{3}{1120 \pi}\, H_I\, m_\phi\, \Gamma_\phi^2\, \frac{a_I}{\ard}\,.
	\end{equation}
	Finally, the differential GW spectrum can be approximated as 
	\begin{equation}
		\ogw h^2(f) \simeq 1.32 \times 10^{-29} \left(\frac{H_I}{2.0 \times 10^{-5}~M_P}\right)^\frac13 \left(\frac{\mu}{10^{12}~\text{GeV}} \right)^\frac{13}{3} \left(\frac{10^{13}~\text{GeV}}{m_\phi} \right)^\frac{25}{6} \left(\frac{f}{10^{7}~\text{Hz}} \right)^3
	\end{equation}
	for $f_\text{min}/2 \lesssim f < f_\text{max}/2$. This contribution is lower than the spectrum produced from pure inflaton-inflaton annihilation; cf. Eq.~\eqref{eq:ogw_inflaton_inflaton}. The difference arises because the graviton production rate in the former case is smaller than in the latter. In particular, the spectrum exhibits a $f^3$ dependence, similar to the GW spectrum generated by relativistic SM degrees of freedom in a thermal plasma~\cite{Ghiglieri:2015nfa, Ghiglieri:2020mhm, Ringwald:2020ist, Bernal:2024jim}. Before closing this section, we note that gravitons can also be produced from the decay of the inflaton into SM particles. However, the corresponding production rate is suppressed compared to that from direct inflaton decay, since the typical energy of the reheaton is lower than the inflaton mass. 
	
	\subsection{Comparison}
	\begin{figure}[t!]
		\def\sepf{0.47}
		\centering
		\includegraphics[scale=\sepf]{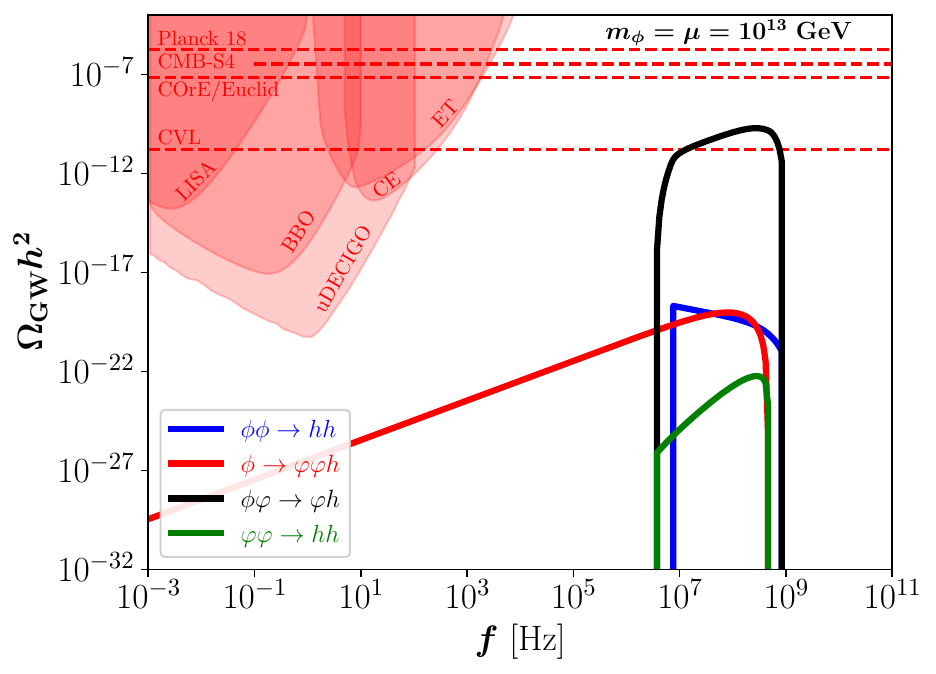}
		\includegraphics[scale=\sepf]{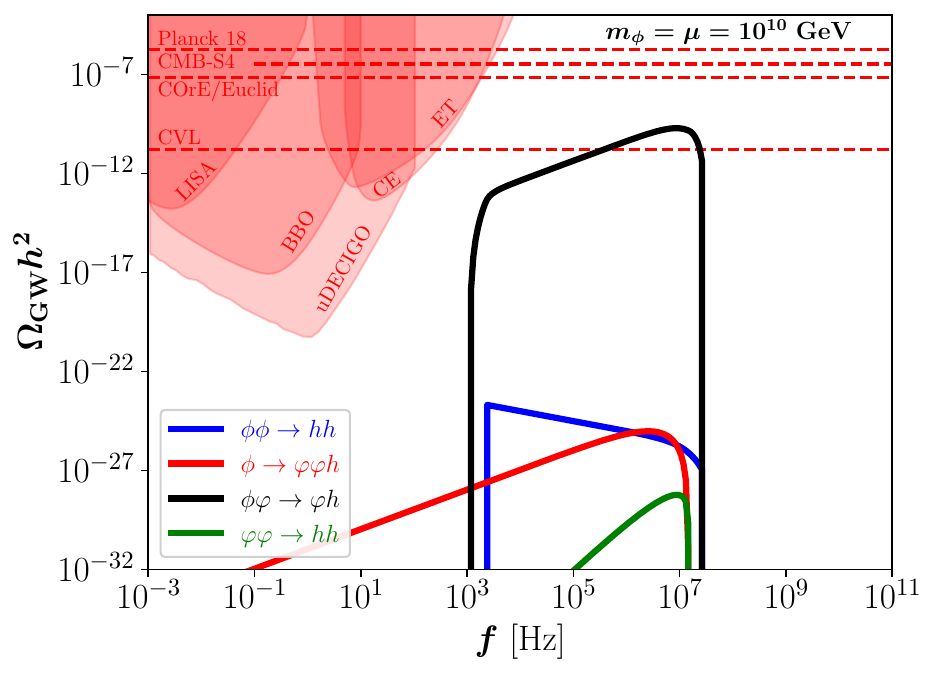}
		\includegraphics[scale=\sepf]{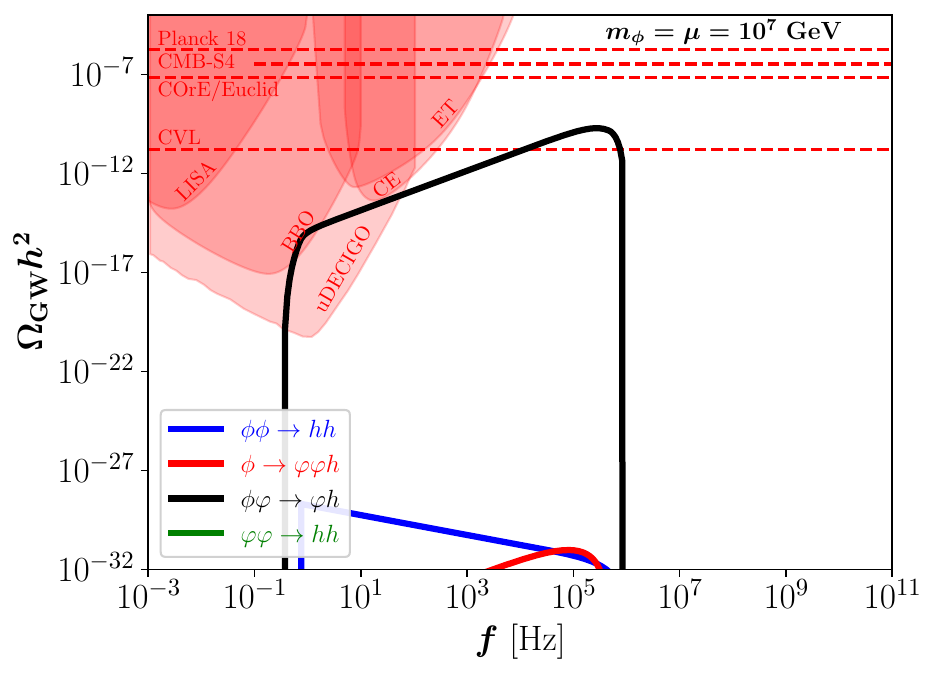}
		\includegraphics[scale=\sepf]{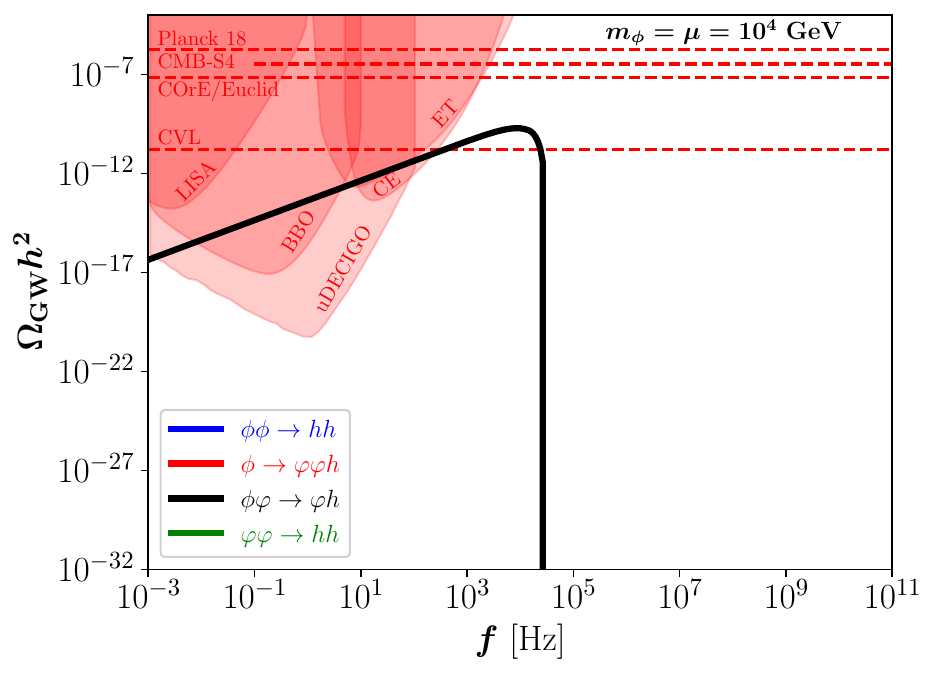}
		\caption{Comparison of the different sources of GWs, for $m_\phi = 10^{13}$~GeV, $10^{10}$~GeV, $10^7$~GeV and $10^4$~GeV, assuming $H_I = 2.0 \times 10^{-5}~M_P$ and $m_\phi = \mu$.} 
		\label{fig:GWall}
	\end{figure} 
	A comparison of the different GW spectra is presented in Fig.~\ref{fig:GWall}, for $m_\phi = 10^{13}$~GeV, $10^{10}$~GeV, $10^7$~GeV and $10^4$~GeV, assuming $H_I = 2.0 \times 10^{-5}~M_P$, and $\mu = m_\phi$, close to the unitarity bound.  The four GW spectra of the tree-level processes with at most four external particles described previously are presented separately: $\phi\, \phi \to h\, h$ (blue), $\phi \to \varphi\, \varphi\, h$ (red), $\phi\, \varphi \to \varphi\, h$ (black), and $\varphi\, \varphi \to h\, h$ (green).
	
	In Fig.~\ref{fig:GWall}, we also present the sensitivity curves of several proposed GW detectors, including the Laser Interferometer Space Antenna (LISA)~\cite{LISA:2017pwj}, the Einstein Telescope (ET)~\cite{Punturo:2010zz, Hild:2010id, Sathyaprakash:2012jk, ET:2019dnz}, the Cosmic Explorer (CE)~\cite{Reitze:2019iox}, the Big Bang Observer (BBO)~\cite{Crowder:2005nr, Corbin:2005ny, Harry:2006fi}, and the ultimate DECIGO (uDECIGO)~\cite{Seto:2001qf, Kudoh:2005as}. The energy stored in GWs behaves similarly to dark radiation, contributing to the effective number of neutrino species, $N_\text{eff}$.\footnote{For the constraint from $N_{\text{eff}}$, we use $\ogw h^2 \lesssim 5.6 \times 10^{-6} \DNeff$, which applies to the integrated energy density over logarithmic frequency~\cite{Caprini:2018mtu}.}  To illustrate these constraints, we include horizontal dashed red lines in Fig.~\ref{fig:GWall}. The Planck 2018 mission provides a 95\% CL measurement of $N_{\text{eff}} = 2.99 \pm 0.34$~\cite{Planck:2018vyg}. Future experiments, such as COrE~\cite{COrE:2011bfs} and Euclid~\cite{EUCLID:2011zbd}, are expected to improve these constraints to $\DNeff \lesssim 0.013$ at the $2\sigma$ level. The next-generation CMB experiment, CMB-S4, is projected to achieve a sensitivity of $\DNeff \lesssim 0.06$~\cite{Abazajian:2019eic}. Additionally, we include a limit of $\DNeff \lesssim 3 \times 10^{-6}$, reported in Ref.~\cite{Ben-Dayan:2019gll}, based on a hypothetical cosmic-variance-limited (CVL) CMB polarization experiment.
	
	Concerning Fig.~\ref{fig:GWall}, we start with the upper left panel, where $m_\phi = \mu = 10^{13}$~GeV. Some comments are in order: $i)$ We find that GWs from the channel $\phi\, \varphi \to \varphi\, h$ exhibit the highest amplitude and have a linear dependency on $f$, which can reach $\ogw \sim \mathcal{O}(10^{-11})$, within the sensitivity of a CVL experiment. $ii)$ The spectra from $\phi\, \varphi \to \varphi\, h$ and $\varphi\, \varphi \to h\, h$ are similar; however, double graviton production has a smaller amplitude due to the additional $1/M_P^2$ suppression. $iii)$ The non-thermal nature of the reheaton (that is, a very large phase-space distribution over a narrow energy band) renders the spectrum from $\phi\, \varphi \to \varphi\, h$ to higher amplitudes and to a smaller frequency range with respect to the Bremsstrahlung $\phi \to \varphi\, \varphi\, h$, even if they share the same matrix element. $iv)$ The GW spectrum from $\phi\, \phi \to h\, h$ annihilation is larger than that from $\varphi\, \varphi \to h\, h$ scatterings. The former follows an $f^{-1/2}$ scaling, mainly due to the non-relativistic nature of the inflaton, whereas the latter exhibits an $f^3$ behavior, similar to GWs generated from relativistic degrees of freedom in the SM thermal bath. 
	
	A decrease in the inflaton mass reduces the energy of the gravitons produced, and therefore the GW spectra move to lower frequencies. In addition, a decrease in $\mu$ translates into a longer reheating era dominated by reheatons (cf. Eq.~\eqref{eq:ardoaI}) and, in turn, a broad range of frequencies for $\phi\, \phi \to h\, h$,  $\phi\, \varphi \to \varphi\, h$ and $\varphi\, \varphi \to h\, h$. For $\mu = m_\phi$, all spectra shift to lower frequencies. However, amplitudes are also reduced, as shown in the upper right and lower panels of Fig.~\ref{fig:GWall}. Interestingly, in the particular case where $\mu \propto m_\phi$, the maximal amplitude of the spectrum from $\phi\, \varphi \to \varphi\, h$ remains unchanged, since it scales as $(\mu/m_\phi)^4$; cf. Eq.~\eqref{eq:GWphivarphi}. It is remarkable that for sufficiently light inflaton masses ($m_\phi \lesssim 10^7$~GeV), the spectrum from $\phi\, \varphi \to \varphi\, h$ falls within the sensitivity of future GW experiments, such as ET, CE, uDECIGO and BBO.
	
\begin{figure}[t!]
		\def\sepf{0.47}
		\centering
		\includegraphics[scale=\sepf]{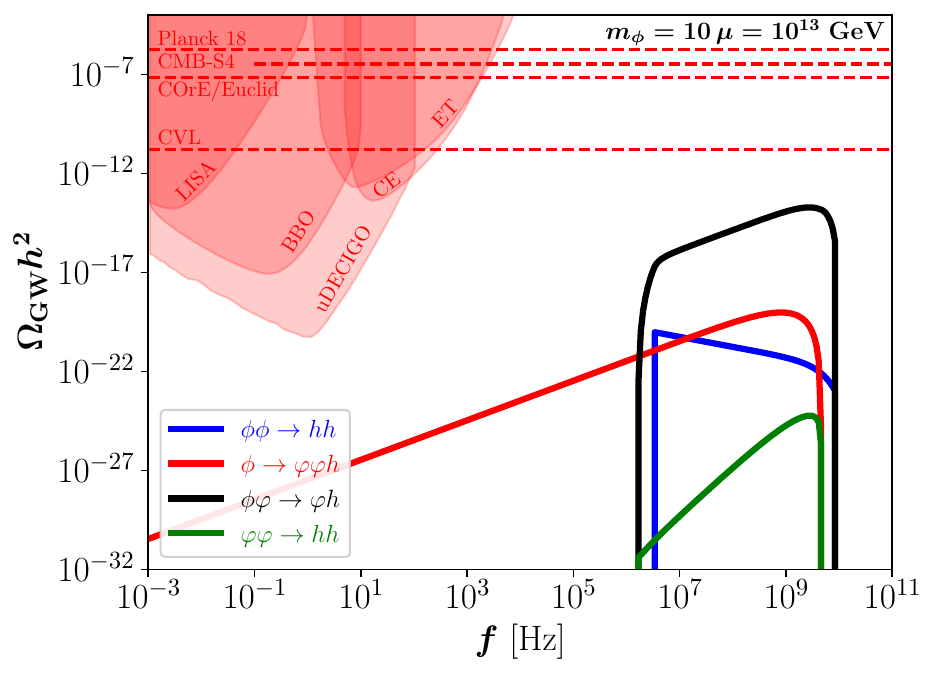}
		\includegraphics[scale=\sepf]{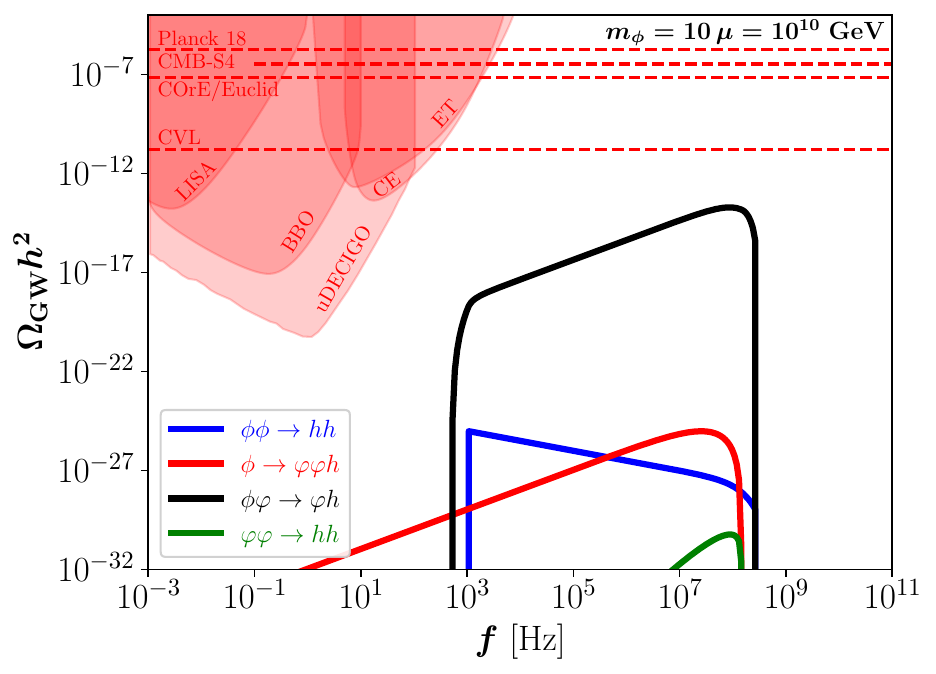}
		\includegraphics[scale=\sepf]{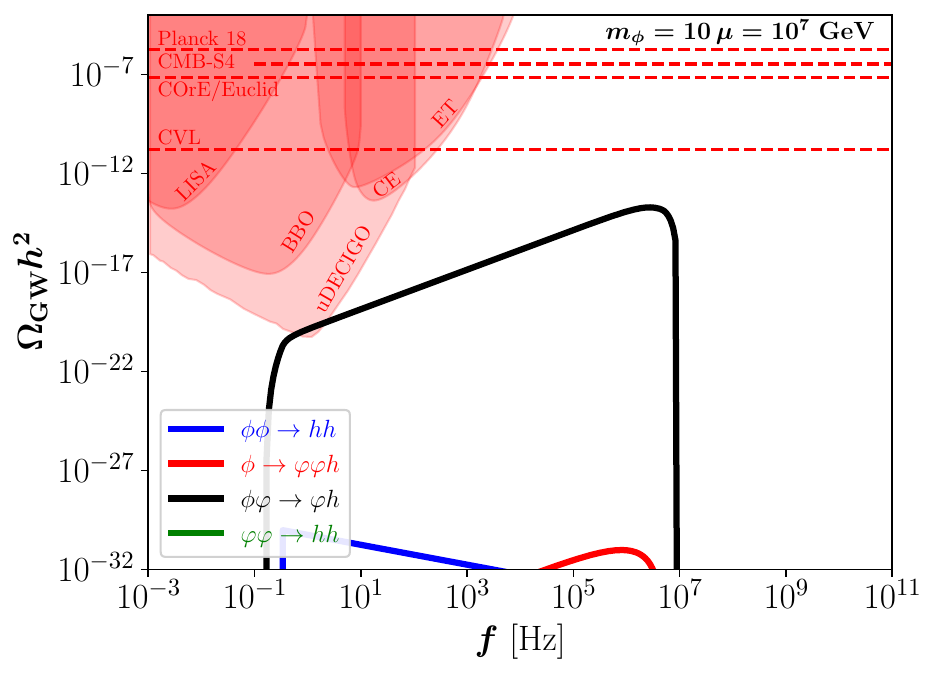}
		\includegraphics[scale=\sepf]{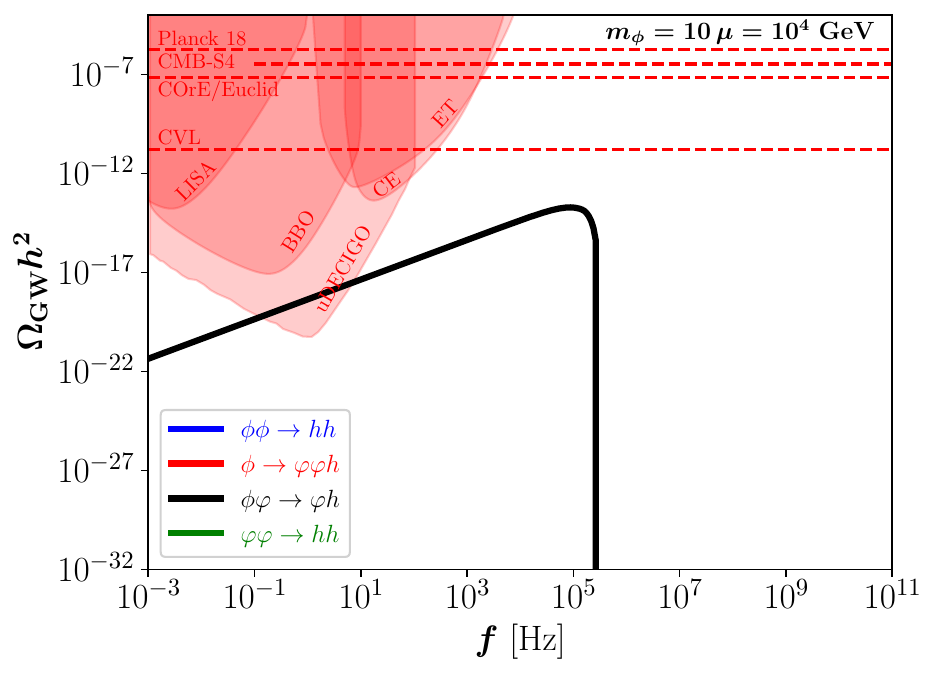}
		\caption{Comparison of the different sources of GWs. Equivalent of Fig.~\ref{fig:GWall} but for $\mu = m_\phi/10$.}
		\label{fig:GWall-10}
	\end{figure}

	In Fig.~\ref{fig:GWall-10}, we consider a scenario with a smaller trilinear coupling $\mu$ with respect to the inflaton mass: $\mu = m_\phi/10$. As expected, the amplitudes of all spectra reduce, and the frequencies span over a larger range. Similarly to the previous case, the GWs produced from $\phi\, \varphi \to \varphi\, h$ have the largest amplitude. In this scenario, the GW spectrum could fall within the sensitivity of uDECIGO for a light inflaton with mass $m_\phi \lesssim 10^7$~GeV.
	
	Before closing, we comment on the effect of $H_I$. For the GW spectrum from $\phi\, \varphi \to \varphi\, h$, the peak is determined by $m_\phi$ and $\mu$ (cf. Eq.~\eqref{eq:GWphivarphi}) and is independent of the inflationary scale $H_I$. Instead, $H_I$ governs the frequency band range, as shown in Eq.~\eqref{eq:frequency_band}. A smaller $H_I$ shifts the lower limit $f_\text{min}$ upward, narrowing the frequency range.
	
	\section{Conclusions} \label{sec:conclusion}
	In this work, a novel gravitational wave (GW) production mechanism is explored for the first time, where gravitons are generated during the {\it non-thermal} phase of the cosmic reheating era. We consider a scenario in which reheating occurs via inflaton decays into a weakly coupled intermediate state (i.e. the reheatons), which subsequently transfer their energy to Standard Model (SM) degrees of freedom, leading to the formation of a thermal SM bath. Before thermalization is complete, a large population of highly energetic reheatons emerges, with energies comparable to the inflaton mass. During this phase, GWs are generated through multiple production channels: $i)$ pure inflaton-inflaton annihilation, $ii)$ graviton Bremsstrahlung from inflaton decays to reheatons, $iii)$ scatterings between an inflaton and a reheaton, and $iv)$ scatterings among reheatons.
	
	We systematically analyze each channel and determine the resulting GW spectrum by solving the Boltzmann equation at the graviton phase-space distribution level. This approach, which has not been previously applied in this context, provides a robust framework that reproduces known results for inflaton-inflaton annihilation and graviton Bremsstrahlung. More notably, we find that the third channel (that is, the reheaton-catalyzed inflaton-to-graviton conversion) leads to a significantly larger amplitude of the GW signal because of the abundance of hard, {\it non-thermal} reheatons in the initial state. Our main results, summarized in Figs.~\ref{fig:GWall} and~\ref{fig:GWall-10}, compare the contributions of different GW production mechanisms. Interestingly, in scenarios with a low inflaton mass ($m_\phi \lesssim 10^7$~GeV), the GW spectrum could lie within the sensitivity range of future experiments such as the Einstein Telescope (ET), Cosmic Explorer (CE), Big Bang Observer (BBO), and ultimate DECIGO (uDECIGO).
	
	In summary, we have identified a novel perturbative source of GWs arising during the pre-thermalization phase of cosmic reheating. The resulting GW signal could be probed by future experiments.
	
	\acknowledgments
	NB acknowledges the Institute of High Energy Physics and the Chinese Academy of Sciences for their hospitality. NB received funding from the grants PID2023-151418NB-I00 funded by MCIU/AEI/10.13039 /501100011033/ FEDER and PID2022-139841NB-I00 of MICIU/AEI/10.13039/501100011033 and FEDER, UE. The work of XJX is supported in part by the National Natural Science Foundation of China (NSFC) under grant No.~12141501 and also by the CAS Project for Young Scientists in Basic Research (YSBR-099).
	YX has received support from the Cluster of Excellence ``Precision Physics, Fundamental Interactions, and Structure of Matter'' (PRISMA$^+$ EXC 2118/1) funded by the Deutsche Forschungsgemeinschaft (DFG, German Research Foundation) within the German Excellence Strategy (Project No. 390831469). This work was performed in part at Aspen Center for Physics, which is supported by National Science Foundation grant PHY-2210452.
	
	\appendix
	
	\section{\boldmath Calculation of \texorpdfstring{$\Msqr{}$}{|M|\^{}2}} \label{sec:M2}
	In this appendix, we present squared amplitudes for several processes discussed in the main text. For reproducibility, we provide the \textsc{Wolfram Mathematica} notebooks that include the derivations of the Feynman rules for the vertices and the calculations of the squared amplitudes, which are available on \href{https://github.com/Fenyutanchan/amplitude-for-inflaton-reheaton-graviton}{\faGithub}.\footnote{\url{https://github.com/Fenyutanchan/amplitude-for-inflaton-reheaton-graviton}}
	
	\subsection{Gravitational Polarization Summation}
	We work in the transverse and traceless gauge with $\omega_\mu\epsilon^{\mu \nu}=0$ and $\eta_{\mu \nu}\epsilon^{\mu \nu} =\epsilon^\mu_\mu=0$, where $\omega_\mu$ denotes the four-momentum of graviton and $\epsilon^{\mu \nu}$ denotes its polarization tensor. The polarization sum reads
	\begin{equation} \label{eq:grapol}
		\sum_\text{pol} \epsilon^{\star\mu\nu} \epsilon^{\alpha\beta} = \frac12 \left(\hat{\eta}^{\mu\alpha} \hat{\eta}^{\nu\beta} + \hat{\eta}^{\mu\beta} \hat{\eta}^{\nu\alpha} - \hat{\eta}^{\mu\nu} \hat{\eta}^{\alpha \beta}\right),
	\end{equation}
	with
	\begin{equation}
		\hat{\eta}_{\mu \nu} \equiv \eta_{\mu \nu} - \frac{\omega_\mu \,\overline{\omega}_{\nu} +\overline{\omega}_{ \mu}\, \omega_{\nu}}{\omega \cdot \overline{\omega}}\,,
	\end{equation}
	where $\omega= (E_\omega, \vec{\omega})$ and $\overline{\omega}= (E_\omega, -\vec{\omega})$.
	
	\subsection{Graviton Production from Scalar Annihilation}
	A pair of gravitons can be generated from the annihilation of a pair of scalars such as inflatons or reheatons, as already shown in Fig.~\ref{fig:Feyn-all}. We label the momenta of the two initial scalars by $p_1$ and $p_2$, and the momenta of the two final gravitons by $p_3$  and $p_4$, respectively. Therefore, the Mandelstam variables are given by $s= (p_1+p_2)^2$, $t= (p_1-p_3)^2$, and $u= (p_1-p_4)^2$ with $s+t+u=2\,m^2$, where $m$ denotes the scalar mass, it being $m_{\phi}$ or $m_{\varphi}$. The matrix elements of diagrams I-A to I-D (or IV-A to IV-D) in Fig.~\ref{fig:Feyn-all} read
	\begin{align}
		i\mathcal{M}_t&= \frac{-i}{M_P} (2\, p_{1 \mu}\,p_{1\nu}) \frac{i}{t-m^2}  \frac{-i}{M_P} (2\, p_{2 \lambda}\,p_{2 \kappa})\epsilon^{\star \mu \nu} \epsilon^{\star \lambda \kappa}, \\
		i\mathcal{M}_u&=  \frac{-i}{M_P} (2\, p_{1 \lambda}\,p_{1\kappa}) \frac{i}{u-m^2}  \frac{-i}{M_P} (2\, p_{2 \mu}\,p_{2 \nu})\epsilon^{\star \mu \nu} \epsilon^{\star \lambda \kappa}, \\
		i\mathcal{M}_s&= \frac{2\,i}{M_P} \left[2\,p_{1 \rho}p_{2 \sigma}-\eta_{\rho \sigma}(m^2 +p_1\cdot p_2) \right] \frac{i}{2\,s}\left[\eta_{\rho \alpha}\eta_{\sigma \beta} +\eta_{\rho \beta}\eta_{\sigma \alpha}-\eta_{\alpha \beta}\eta_{\rho \sigma}\right]  V_{\alpha \beta \mu \nu \lambda \kappa}\, \epsilon^{\star \mu \nu} \epsilon^{\star \lambda \kappa}, \\
		i\mathcal{M}_4&= \frac{i}{M^2_P} \left[\eta_{\mu \nu}\eta_{\lambda \kappa} -\eta_{\mu \lambda }\eta_{\nu \kappa} -\eta_{\mu \kappa }\eta_{\nu \lambda}\right] (-p_1\cdot p_2 -m^2) \nonumber\\
		&\qquad +\left[2(\eta_{\nu \lambda}\, p_{2\mu}\,p_{2\kappa}+\eta_{\mu \kappa} p_{2\lambda}\,p_{2\nu}+\eta_{\mu \lambda} \,p_{2\nu}\,p_{2\kappa}+\eta_{\nu \kappa} \,p_{2\lambda}\,p_{2\mu})\right] \epsilon^{\star \mu \nu} \epsilon^{\star \lambda \kappa},
	\end{align}
	where $V_{\alpha \beta \mu \nu \lambda \kappa}$ denotes the triple-graviton vertex. The corresponding Feynman rules can be found in Ref.~\cite{Choi:1994ax}. Note that any term with $p_{3 \mu}$, $p_{3\nu}$, $p_{4\lambda}$, $p_{4\kappa}$, $\eta_{\mu \nu}$, or $\eta_{\lambda \kappa}$ would vanish in the matrix element once combined with the graviton polarization tensors due to the transverse condition and traceless condition. We find
	\begin{align}\label{eq:M}
		|\mathcal {M}_t| ^2 &= \frac{4\Big[m^4 -2m^2 t + t(s+t)\Big]^4}{M_P^4 s^4 (m^2-t)^2} \,, \\
		|\mathcal {M}_u| ^2 &= \frac{4\Big[m^4 -2m^2 u + u(s+u)\Big]^4}{M_P^4 s^4 (m^2-u)^2} \,, \\
		|\mathcal {M}_s| ^2 &=\frac{2\left( s^2 +3st +3t^2\right)^2}{M_P^4 s^2} \,, \\
		|\mathcal {M}_4| ^2 &= \frac{2\left[4m^4-8m^2t+(s+2t)^2\right]^2}{M_P^4\, s^2}\,,\\
		\mathcal{M}_s \mathcal{M}^{\star}_t&= \frac{-2\left[m^4 -2m^2t+t(s+t)\right]^2 \left[3m^4 -m^2(s+6t)+s^2 +3st+3t^2\right] }{M_P^4 s^3(m^2-t)}\,,\\
		\mathcal{M}_s \mathcal{M}^{\star}_u&= \frac{-2\left[m^4 -2m^2t+t(s+t)\right]^2 \left[3m^4 -m^2(s+6t)+s^2 +3st+3t^2\right] }{M_P^4 s^3(s+t-m^2)}\,,\\
		\mathcal{M}_s \mathcal{M}^{\star}_4& = \frac{-2\left[4m^4 -8m^2t+(s+2t)^2\right] \left[3m^4 -m^2(s+6t)+s^2 +3st+3t^2\right] }{M_P^4 s^2}\,,\\
		\mathcal{M}_t \mathcal{M}^{\star}_u&= \frac{4\left[m^4 -2m^2 t+t(s+t)\right]^4}{M_P^4 s^4 (m^2-t)(s+t-m^2)}\,,\\
		\mathcal{M}_t \mathcal{M}^{\star}_4&= \frac{2\left[m^4 -2m^2 t+t(s+t)\right]^2 \left[4m^4 -8m^2t+(s+2t)^2\right]}{M_P^4 s^3 (m^2-t)}\,,\\
		\mathcal{M}_u \mathcal{M}^{\star}_4&= \frac{2\left[m^4 -2m^2 t+t(s+t)\right]^2 \left[4m^4 -8m^2t+(s+2t)^2\right]}{M_P^4 s^3 (s+t-m^2)}\,.
	\end{align}
	The total squared matrix elements are
	\begin{align} \label{eq:M2-anni}
		|\mathcal{M}|^2 &= |\mathcal{M}_s+ \mathcal{M}_t+\mathcal{M}_u+\mathcal{M}_4|^2 \nonumber \\
		& = |\mathcal{M}_s|^2+|\mathcal{M}_t|^2+|\mathcal{M}_u|^2+|\mathcal{M}_4|^2 \nonumber\\
		& \quad + 2\left(\mathcal{M}_s \mathcal{M}^{\star}_t+\mathcal{M}_s \mathcal{M}^{\star}_u+\mathcal{M}_s \mathcal{M}^{\star}_4+\mathcal{M}_t \mathcal{M}^{\star}_u+\mathcal{M}_t \mathcal{M}^{\star}_4+\mathcal{M}_u \mathcal{M}^{\star}_4\right) \nonumber\\
		& = \frac{2}{s^2 \left(t - m^2\right)^2 \left(s + t - m^2\right)^2 M_P^4}   \nonumber\\
		& \quad \times \Big[m^{16} - 8m^{14} t + 4\, m^{12} t (s + 7t) - 8\, m^{10} t^2 (3 s+ 7 t) \nonumber\\
		&\qquad  + m^8 \left(s^4 + 6 s^2 t^2 + 60 s t^3 + 70 t^4\right) - 8 m^6 t^3 (s + t) (3 s +7 t)\nonumber\\
		& \qquad + 4 m^4\, t^3 (s + t)^2 (s + 7 t) - 8 m^2 t^4(s + t)^3 +t^4 (s + t)^4 \Big].
	\end{align}
	We notice that Eq.~\eqref{eq:M2-anni} is more general compared to the existing ones in the literature. For non-relativistic inflatons annihilating into gravitons, we have
	\begin{equation} \label{eq:M2-anni-nr}
		\Msqr{\phi\phi\to hh} = \frac{2\, m_\phi^4}{M_P^4}\,,
	\end{equation}
	thus reproducing the result presented in Ref.~\cite{Choi:2024ilx}. 
	For relativistic reheatons annihilating into gravitons, Eq.~\eqref{eq:M2-anni} reduces to 
	\begin{equation} 
		\Msqr{\varphi\varphi\to hh} = \frac{2\, t^2\, (s + t)^2}{M_P^4\, s^2}\,,
	\end{equation}
	which agrees with Eq.~(25) in Ref.~\cite{Ghiglieri:2022rfp}.
	
	\subsection{Graviton Production from Bremsstrahlung}\label{sec:app1to3}
	For momenta labeled as $\phi(l) \to \varphi(p)\, \varphi(q)\, h_{\mu \nu} (\omega)$, the matrix elements are~\cite{Barman:2023ymn}
	\begin{align}
		i\mathcal{M}_1 &= -\frac{i\, \mu}{M_P}\, \frac{l_\mu\, l_\nu\, \epsilon^{\star\mu\nu}}{l\cdot \omega}\,, \\
		i\mathcal{M}_2 &= +\frac{i\, \mu}{M_P}\, \frac{p_\mu\, p_\nu\, \epsilon^{\star\mu\nu}}{p \cdot \omega}\,, \\
		i\mathcal{M}_3 &= +\frac{i\, \mu}{M_P}\, \frac{q_\mu\, q_\nu\, \epsilon^{\star\mu\nu}}{q\cdot \omega}\,,\\
		i\mathcal{M}_4 &\propto \eta_{\mu \nu} \epsilon^{\mu \nu}  =0\,,
	\end{align} 
	for the initial-sate graviton emitted from the inflaton ($\mathcal{M}_1$), the final-state reheatons ($\mathcal{M}_2$ and $\mathcal{M}_3$) and the vertex ($\mathcal{M}_4$), respectively. The last matrix element $\mathcal{M}_4$ vanishes due to the traceless condition of the graviton polarization tensor. Moreover, the matrix element $\mathcal{M}_1$ vanishes since the inflaton is at rest, that is, $l = (m_\phi, \vec 0)$. This, in other words, implies that the source of GW in this case is $T^{ij}_\phi =0$, as commented just below Eq.~\eqref{eq:Tmunu_inflaton}. Note that $T^{ij}$ in momentum space corresponds to the three-momentum. Another way to see this is to directly compute the squared matrix element $|\mathcal{M}_1|^2$, and we find 
	\begin{equation}\label{eq:M1_phivarphi}
		|\mathcal{M}_1|^2 = \frac{\mu^2 \left[l^2 (\omega \cdot \overline{\omega}) -2 (l\cdot \omega) (l\cdot \overline{\omega}) \right]^2}{2 M_P^2 (l\cdot \omega)^2 (\omega \cdot \overline{\omega})^2}\,,
	\end{equation}
	which is $ \frac{\mu^2 \left[2m_\phi^2 E_\omega^2-2m_\phi^2 E_\omega^2 \right]^2}{8\, M_P^2 E_\omega^2 m_\phi^2 E_\omega^4 } =0$ in the non-relativistic limit of the inflaton. Note that $\mathcal{M}_2$ and $\mathcal{M}_3$ do not vanish because the produced reheatons $\varphi$ are not at rest. An equivalent argument would be the source of GW $T^{ij}_\varphi \neq 0$.  
	
	Using the conservation of four momentum $l=p+q+\omega$ and the transverse condition of graviton, we notice that  the third diagram can be smartly rewritten as 
	\begin{equation}
		i\mathcal{M}_3 = \frac{+i\, \mu}{M_P}\, \frac{p_\mu\, p_\nu\, \epsilon^{\star\mu\nu}}{(l-p)\cdot \omega}\,,
	\end{equation}
	which greatly simplifies the calculation. We find
	\begin{align}
		|\mathcal{M}_2|^2 &  = \frac{\mu^2 \left[p^2 (\omega \cdot \overline{\omega}) -2 (p\cdot \omega) (p\cdot \overline{\omega}) \right]^2}{2 M_P^2 (p\cdot \omega)^2 (\omega \cdot \overline{\omega})^2}\,, \\
		|\mathcal{M}_3|^2 &= \frac{\mu^2 \left[p^2 (\omega \cdot \overline{\omega}) -2 (p\cdot \omega) (p\cdot \overline{\omega}) \right]^2}{2 M_P^2 (l\cdot \omega - p\cdot \omega)^2 (\omega \cdot \overline{\omega})^2}\,, \\
		\mathcal{M}_2\mathcal{M}^{\star}_3 &= \frac{\mu^2 \left[p^2 (\omega \cdot \overline{\omega}) -2 (p\cdot \omega) (p\cdot \overline{\omega}) \right]^2}{2 M_P^2 (p\cdot \omega)(l\cdot \omega - p\cdot \omega)  (\omega \cdot \overline{\omega})^2}\,,
	\end{align}
	and the total squared matrix element is 
	\begin{align} \label{eq:1to3}
		|\mathcal{M}_{\phi \to \varphi \varphi h}|^2 &= |\mathcal{M}_2|^2 +|\mathcal{M}_3|^2 + 2\mathcal{M}_2\mathcal{M}^{\star}_3 =\frac{\mu^2}{2M_P^2} \frac{(l\cdot \omega)^2 \left[p^2  (\omega \cdot \overline{\omega}) - 2(p\cdot \omega) (p\cdot \overline{\omega}) \right]^2}{(l \cdot \omega -p \cdot \omega)^2(p\cdot \omega)^2 (\omega \cdot \overline{\omega})^2 }\nonumber \\
		&\simeq \frac{2\mu^2}{M_P^2} \frac{(l\cdot \omega)^2   (p\cdot \overline{\omega})^2}{(l \cdot \omega -p \cdot \omega)^2(\omega \cdot \overline{\omega})^2 } =\frac{2\mu^2}{M_P^2} \frac{(l\cdot \omega)^2   (p\cdot \overline{\omega})^2}{(q \cdot \omega)^2(\omega \cdot \overline{\omega})^2 } \,,
	\end{align}
	where in the last second step we have taken $\mr \to 0$. We also note that our result is invariant under $p \leftrightarrow q$ as expected. 
	
	Without loss of generality, we take $\omega^{\mu}=(E_{h},E_{h},0,0)$, $\overline{\omega}^{\mu}=(E_{h},-E_{h},0,0)$, $l^{\mu}=(m_{\phi},0,0,0)$, and $p^{\mu}=(E_{p},\ p_{x},\ p_{y},\ p_{z})$. Then using the on-shell conditions of $p^{\mu}$ and $q^{\mu}$ ($p\cdot p=q\cdot q=0$) as well as $l\cdot p=m_{\phi}E_{p}$ and $l\cdot\omega=m_{\phi}E_{h}$, it is straightforward to express $E_{h}$, $E_{p}$, $p_{x}$, and $p_{y}^{2}+p_{z}^{2}$ in terms of $l\cdot p$, $l\cdot\omega$, $p\cdot p$, and $q\cdot q$.  This allows us to further obtain
	\begin{equation}
		\frac{p\cdot\overline{\omega}}{q\cdot\omega}=1-2\frac{l\cdot\omega}{m_{\phi}^{2}}=1-2\frac{E_{h}}{m_{\phi}}\,,\label{eq:p-q-sym}
	\end{equation}
	and simplify Eq.~\eqref{eq:1to3} to 
	\begin{equation} \label{eq:M2-decay-final}
		|\mathcal{M}_{\phi\to\varphi\varphi h}|^{2} = \frac{2\, \mu^{2}}{M_{P}^{2}} \left(1-\frac{m_{\phi}}{2\, E_{h}}\right)^{2}.
	\end{equation}
	
	\subsection{Graviton Production from Inflaton and Reheaton Scattering} \label{sec:phi_varphi}
	For momenta labeled as $\phi(l)\, \varphi(q) \to \varphi(p)\, h_{\mu \nu}(\omega)$, and the matrix elements are~\cite{Xu:2024fjl}
	\begin{align} 
		i\mathcal{M}_u &= -\frac{i\, \mu}{M_P}\, \frac{l_\mu\, l_\nu\, \epsilon^{\star\mu\nu}}{l\cdot \omega}\,, \\
		i\mathcal{M}_s &= +\frac{i\, \mu}{M_P}\, \frac{p_\mu\, p_\nu\, \epsilon^{\star\mu\nu}}{p \cdot \omega}\,, \\
		i\mathcal{M}_t &= -\frac{i\, \mu}{M_P}\, \frac{q_\mu\, q_\nu\, \epsilon^{\star\mu\nu}}{q\cdot \omega}\,,\\
		i\mathcal{M}_4 &\propto \eta_{\mu \nu} \epsilon^{\mu \nu} =0\,,
	\end{align} 
	where the last matrix element $\mathcal{M}_4 $ vanishes due to the traceless condition for graviton polarization tensor. We note if changing the reheaton $\varphi$ from the initial state to the final state, we reproduce the case $1\to 3$ in the previous subsection. Consequently, the matrix elements for these $2\to 2$ scatterings are similar to those presented in the previous case of $1 \to 3$ Bremsstrahlung, except that there is an extra minus sign in $\mathcal{M}_t$ compared to $\mathcal{M}_3$.
	
	We define Mandelstam variables as $s=(l+q)^2$, $t=(l-p)^2$, and $u = (l-\omega)^2$ with $s+t+u =2m_{\varphi}^2 +m_\phi^2$. The general total squared matrix element is shown to be:
	\begin{equation} \label{eq:Mphivarphi}
		|\mathcal{M}_{\phi \varphi \to \varphi h}|^2 = \frac{2\mu^2 \left[2m_{\varphi}^6 -m_{\varphi}^4 (m_\phi^2+s+t) +m_{\varphi}^2(m_\phi^2(s+t)-2st)+st(s+t-m_\phi^2)\right]^2}{M_P^2(s-m_{\varphi}^2)^2 (t-m_{\varphi}^2)^2 (s+t-2m_{\varphi}^2)^2}.
	\end{equation}
	In the limit of ultra-relativistic reheatons ($m_{\varphi} \to 0$), it reduces to
	\begin{equation}
		|\mathcal{M}_{\phi \varphi \to \varphi h}|^2 = \frac{ 2\mu^2 (s+t-m_\phi^2)^2}{M_P^2 (s+t)^2}\,.
	\end{equation}
	Additionally, if one further demands non-relativistic inflatons, $l=(m_\phi, 0, 0, 0)$, we have $u = (l - \omega)^2 = m_{\phi}^{2} - 2 l\cdot\omega = m_{\phi}^{2} - 2 m_{\phi} E_{h}$. Then using $s+t=m_{\phi}^{2}-u$, we can simply Eq.~\eqref{eq:Mphivarphi} as
	\begin{equation} \label{eq:M2-coanni-final}
		|\mathcal{M}_{\phi\varphi\to\varphi h}|^{2}=\frac{\mu^{2}}{2\,M_{P}^{2}}\left(2-\frac{m_{\phi}}{E_{h}}\right)^{2},
	\end{equation}
	which matches Eq.~\eqref{eq:M2-decay-final} and agrees with Ref.~\cite{Xu:2024fjl}.
	
	Finally, we note here that $\mathcal{M}_u$ vanishes for non-relativistic $\phi$.  This can also be easily seen by noticing that
	\begin{equation} \label{eq:phivarphiMu}
		|\mathcal{M}_u|^2 = \frac{2\mu^2 \left[l^2 (\omega \cdot \overline{\omega}) -2 (l\cdot \omega) (l\cdot \overline{\omega}) \right]^2}{M_P^2 (u-m_\phi^2)^2 (\omega \cdot \overline{\omega})^2} = \frac{2\mu^2 \left[2m_\phi^2 E_h^2-2m_\phi^2 E_h^2 \right]^2}{16\, M_P^2 E_h^2 m_\phi^2 E_h^4 } =0\,,
	\end{equation}
	for the same reason that Eq.~\eqref{eq:M1_phivarphi} vanishes.
	
	\section{\boldmath Collision Terms \texorpdfstring{$\Gamma_h$}{Γh}} \label{sec:collision}
	In this section, we present in great detail the computation of the collision terms for the graviton production for the processes $\phi \to \varphi\, \varphi\, h$ and $\phi\, \varphi \to \varphi\, h$, which are substantially more involved than $\phi\, \phi \to h\, h$ and $\varphi\, \varphi \to h\, h$.
	
	\subsection[\texorpdfstring{$\phi \to \varphi\, \varphi\, h$}{ϕ → φ φ h}]{\boldmath $\phi \to \varphi\, \varphi\, h$} \label{sec:Gh2}
	Labeling the particles in $\phi \to \varphi\, \varphi\, h$ from the left to the right by numbers 1 to 4 (with $E_4 = E_h$), $\Gamma_h$ is then given by
	\begin{align}
		\Gamma_{h} & =\frac{1}{2\, E_h}\int d\Pi_{1}\, d\Pi_{2}\, d\Pi_{3}\, f_{1}\, \frac{\Msqr{\phi \to \varphi\varphi h}}{2}\, (2\pi)^4\, \delta^{(4)}(p_1 - p_2 - p_3 - p_4)\nonumber \\
		& =\frac{\Msqr{\phi \to \varphi\varphi h}}{2\, E_h} \int d\Pi_{1}\, d\Pi_{2}\, f_{1}\, \frac{1}{2E_{3}}\, 2\pi\, \delta\left(E_{1}-E_{2}-E_{3}-E_h\right),\label{eq:x-1}
	\end{align}
	where $E_{3} = \sqrt{E_{2} + E_{4} + 2 c_{24}\, E_{2}\, E_h}$ with $c_{24} \equiv \vec{p}_{2} \cdot \vec{p}_{4}/(E_{2}\, E_h)$ being the cosine of the angle between $\vec{p}_{2}$ and $\vec{p}_{4}$, and noticing that $\Msqr{\phi\to\varphi\varphi h}$ can be moved outside the phase-space integral since it depends only on $E_h$; cf Eq.~\eqref{eq:M2-phi-decay}. To integrate out the delta function, we should view it as a function of $E_{2}$, $E_h$, and $c_{24}$:
	\begin{equation}
		\delta\left(E_{1} - E_{2} - E_{3} -E_h\right) = \delta\left[F(E_{2}, E_h, c_{24})\right]\label{eq:x-2}
	\end{equation}
	with $F(E_{2}, E_h, c_{24}) = E_{1} - E_{2} - E_h - \sqrt{E_{2}^2 + E_h^2 + 2c_{24}\, E_{2}\, E_h}$.
	It is straightforward to check that $F$ as a function of $c_{24}$ only has one root within $c_{24}\in[-1,\ 1]$, if and only if $E_h$ and $E_{2}$ are in a triangle bounded by 
	\begin{equation}
		E_h < \frac{m_\phi}{2}\,,\quad \ E_{2}<\frac{m_\phi}{2}\,, \quad \ E_{2}+E_h>\frac{m_\phi}{2}\,.\label{eq:triangle-1}
	\end{equation}
	With this triangle, we integrate out the delta function with $c_{24}$ and then integrate $E_{2}$ from $m_\phi/2 - E_h$ to $m_\phi/2$, arriving at
	\begin{equation} \label{eq:x-3}
		\Gamma_h = \frac{\Msqr{\phi \to \varphi\varphi h}}{64 \pi}\, \frac{n_\phi}{m_\phi\, p_h}\,.
	\end{equation}
	
	\subsection[\texorpdfstring{$\phi\, \varphi \to \varphi\, h$}{ϕ φ → φ h}]{\boldmath $\phi\, \varphi \to \varphi\, h$} \label{sec:Gh3}
	The calculation of the collision term for $\phi\varphi\to\varphi h$ is similar to the calculation for $\phi\to\varphi\varphi h$ in the previous subsection; however, there are some differences. First, Eq.~\eqref{eq:x-1} is changed to
	\begin{align} \label{eq:x-1-1}
		\Gamma_{h} & =\frac{1}{2E_h}\int d\Pi_{1}\, d\Pi_{2}\, d\Pi_{3}\, f_{1}\, f_{2}\,  \Msqr{\phi\, \varphi \to \varphi\, h}\, (2\pi)^4\, \delta^{(4)}(p_1 + p_2 - p_3 - p_4)\nonumber \\
		& =\frac{\Msqr{\phi\, \varphi \to \varphi\, h}}{2 E_h}\int d\Pi_{1}\, d\Pi_{2}\, f_{1}\, f_{2}\, \frac{1}{2E_{3}}\, 2\pi\, \delta\left(E_{1}+E_{2}-E_{3}-E_h\right).
	\end{align}
	Following a similar analysis on the delta function, we find that the triangle region bounded by Eq.~\eqref{eq:triangle-1} is changed to a region constrained by
	\begin{equation}
		\frac{m_\phi}{2} < E_h\,, \qquad E_h - E_{2} < \frac{m_\phi}{2}\,,\label{eq:triangle-2}
	\end{equation}
	which for $E_{2}<\frac{m_\phi}{2}$ would imply that $E_h < m_\phi$, as already mentioned above. Then by integrating out $c_{24}$, we obtain
	\begin{equation} \label{eq:x-8}
		\Gamma_{h} = \frac{\Msqr{\phi\, \varphi \to \varphi\, h}\, n_{\phi}}{32\pi\,  m_\phi\, E_h^2} \int_{p_{\varphi}^{\min}}^{p_{\varphi}^{\max}} f_{\varphi}\, dp_{\varphi}\,,
	\end{equation}
	where $p_{\varphi}^{\min} \equiv E_h - \frac{m_{\phi}}{2}$ and $p_{\varphi}^{\max} = \frac{m_{\phi}}{2}$. Here, $p_{\varphi}^{\max } =\frac{m_{\phi}}{2}$ is imposed because $f_{\varphi}$ has a cut-off point at $\frac{m_{\phi}}{2}$. If $f_{\varphi}$ is a thermal distribution, then $p_{\varphi}^{\max} = \infty$ should be used. After substituting Eq.~\eqref{eq:f-reheaton} into Eq.~\eqref{eq:x-8}, the integration can be performed numerically. Analytically, we find that the result can be approximated by 
	\begin{equation} \label{eq:x-9}
		\Gamma_h \simeq \frac{\pi}{2}\, \frac{\mu^2\, n_\phi\, n_\varphi}{M_P^2\, E_h^2\, m_\phi^3} \left(1 - \frac{m_\phi}{2\, E_h}\right)^2\, \tilde\Theta\left[\frac12 < \frac{E_h}{m_\phi} < 1\right].
	\end{equation}
	
	\section{Comparing Analytical with Numerical Results}\label{sec:compare}
	In this section, we first offer some detailed steps for deriving the analytical expression for the gravitational spectra shown in the main text. Then, we compare them with the fully numerical results. The GW spectrum at present is 
	\begin{align} \label{eq:GW_Spectrum}
		\ogw(a_0) & = \frac{1}{\rho_c}\, f_h\left[a_0, p_h(a_0)\right] \frac{p_h^4(a_0)}{\pi^2}  = \ogw(\ard) \left(\frac{\ard}{a_0}\right)^4 \nonumber \\
		&= \frac{1}{\rho_c}\,  f_h\left[\ard, p_h(\ard)\right] \frac{p_h^4(\ard)}{\pi^2} \left(\frac{\ard}{a_0}\right)^{4}.
	\end{align}
	
	We start by presenting the main channel in this work, namely $\phi \varphi \to \varphi h$, where the phase-space distribution function at $\ard$ is given by Eq.~\eqref{eq:fh_phivarphi}. The corresponding GW spectrum is
	\begin{align} \label{eq:GW_phivarphi}
		\ogw^{\phi \varphi \to \varphi h}(f) &= \frac{1}{\rho_c}  \frac{\pi}{4} \frac{M_P^2\, \mu^2\, H_I^2\, \Gamma_\phi}{m_\phi^4} \left(\frac{a_I}{\ard}\right)^3 \frac{2\pi f a_0/\ard}{\pi^2} \left(\frac{\ard}{a_0}\right)^4 = \frac{1}{\rho_c}  \frac12 \frac{M_P^2\, \mu^2\, H_I^2\, \Gamma_\phi}{m_\phi^4} \left(\frac{a_I}{a_0}\right)^3 f \nonumber \\
		&\simeq \frac{1}{\rho_c}  \frac{1}{2} \frac{M_P^2\, \mu^2\, H_I^2\, \Gamma_\phi}{m_\phi^4}  \left( \frac{2 \Gamma_\phi}{3H_I}\right)^{2}   \left[ \frac{\rR (a_0)} {\frac{4}{3} \Gamma_\phi^2 M_P^2} \right]^{3/4} f\nonumber \\
		&\simeq 1.29 \times 10^{-15} \left( \frac{\mu}{10^{12}~\text{GeV}}\right)^{5}  \left( \frac{10^{13}~\text{GeV}}{m_\phi}\right)^{11/2}  \left( \frac{f}{10^{7}~\text{Hz}}\right).
	\end{align}
	We recall that the term $H_I^2$ in the first line of Eq.~\eqref{eq:GW_phivarphi} originates from the number density $n_\phi$ in the initial state of the scattering process under consideration. A larger $H_I$ implies that GW production occurs earlier, leading to a greater redshift for the produced gravitons. As seen in the last second step, the contribution from the inflationary scale $H_I$ ultimately cancels out due to this redshift effect. Consequently, the maximal amplitude for the GW spectrum remains independent of $H_I$.
	
	For the Bremsstrahlung channel $\phi \to \varphi\, \varphi\, h$, the phase-space distribution function at $\ard$ is given by Eq.~\eqref{eq:fh-Brems}, and the corresponding GW spectrum is
	\begin{align} \label{eq:GW_phi}
		\ogw^{\phi \to \varphi \varphi h}(f) &\simeq  \frac{1}{\rho_c} \frac{1}{64 \pi}\, \frac{\mu^2\, H_I}{p_h^3(\ard)} \left(\frac{a_I}{\ard}\right)^\frac32 \frac{p_h^4(\ard)}{\pi^2} \left(\frac{\ard}{a_0}\right)^{4} \nonumber \\
		&=  \frac{1}{\rho_c}   \frac{\mu^2\, H_I}{64 \pi} \left(\frac{a_I}{\ard}\right)^\frac32 \frac{2\pi f a_0/\ard}{\pi^2} \left(\frac{\ard}{a_0}\right)^{4} \nonumber \\
		&\simeq 2.15\times 10^{-20}  \left(\frac{\mu}{10^{12}~\text{GeV}} \right)  \left(\frac{m_\phi}{10^{13}~\text{GeV}} \right)^{1/2} \left(\frac{f}{10^{7}~\text{Hz}} \right).
	\end{align}
	Similarly to the previous case, the dependence of $H_I$ cancels out, leading to a maximal amplitude for the spectrum independent of it.
	
	For $\phi \phi \to hh$, the phase-space distribution function at $\ard$ given in Eq.~\eqref{eq:fh-infinf} generates a GW spectrum
	\begin{align} \label{eq:GW_phiphi}
		\ogw^{\phi \phi \to hh}(f) &\simeq \frac{1}{\rho_c}  \frac{9\pi\, m_\phi^\frac32\, H_I^3} {16} \frac{1}{p_h^\frac92(\ard)}  \left( \frac{a_I}{\ard}\right)^\frac92   \frac{p_h^4(\ard)}{\pi^2}\left(\frac{\ard}{a_0}\right)^{4} \nonumber \\
		&\simeq  \frac{1}{\rho_c}  \frac{9\, m_\phi^\frac32  H_I^3} {16\, \pi} \frac{1}{ \left(2\pi f a_0/\ard \right)^\frac12}  \left( \frac{a_I}{\ard}\right)^\frac92  \left(\frac{\ard}{a_0}\right)^{4} \simeq  \frac{1}{\rho_c}  \frac{9 m_\phi^\frac32  H_I^3} {16\, \pi} \frac{1}{ \left(2\pi f  \right)^\frac12 }  \left( \frac{a_I}{a_0}\right)^\frac92 \nonumber \\
		&\simeq 1.25 \times 10^{-19} \left( \frac{\mu}{10^{12}~\text{GeV}}\right)^\frac32  \left( \frac{m_\phi}{10^{13}~\text{GeV}}\right)^\frac34  \left( \frac{f}{10^{7}~\text{Hz}}\right)^{-\frac12}.
	\end{align}
	
	Finally, for $\varphi \varphi \to hh$, the phase-space distribution function at $\ard$ is given by Eq.~\eqref{eq:fh_varphivarphi}, and the corresponding GW spectrum is
	\begin{align} \label{eq:GW_varphivarphi}
		\ogw^{\varphi \varphi \to hh} &\simeq  \frac{1}{\rho_c}    \frac{9\, \pi\,H_I\, \Gamma^2}{140\, m_\phi^2\, p_h(\ard)} \frac{a_I}{\ard} \frac{p_h^4(\ard)}{\pi^2}  \left(\frac{\ard}{a_0}\right)^{4} \nonumber \\
		&=  \frac{1}{\rho_c}   \frac{9\,H_I \, \Gamma^2\, \pi}{140\, m_\phi^2} \frac{\left[2\pi f a_0/\ard \right]^3}{\pi^2} \frac{a_I}{\ard} \left(\frac{\ard}{a_0}\right)^{4} =  \frac{1}{\rho_c}   \frac{18\, H_I\, \Gamma^2\,\pi^2}{35\, m_\phi^2} \left(\frac{a_I}{a_0} \right) f^3 \nonumber \\
		&\simeq 2.9 \times 10^{-29} \left(\frac{H_I}{2.0 \times 10^{-5}~M_P} \right)^\frac13 \left(\frac{\mu}{10^{12}~\text{GeV}} \right)^\frac{13}{3}  \left(\frac{10^{13}~\text{GeV}}{m_\phi} \right)^\frac{25}{6} \left(\frac{f}{10^{7}~\text{Hz}} \right)^3.
	\end{align}
	
	\begin{figure}[t!]
		\def\sepf{0.6}
		\centering
		\includegraphics[scale=\sepf]{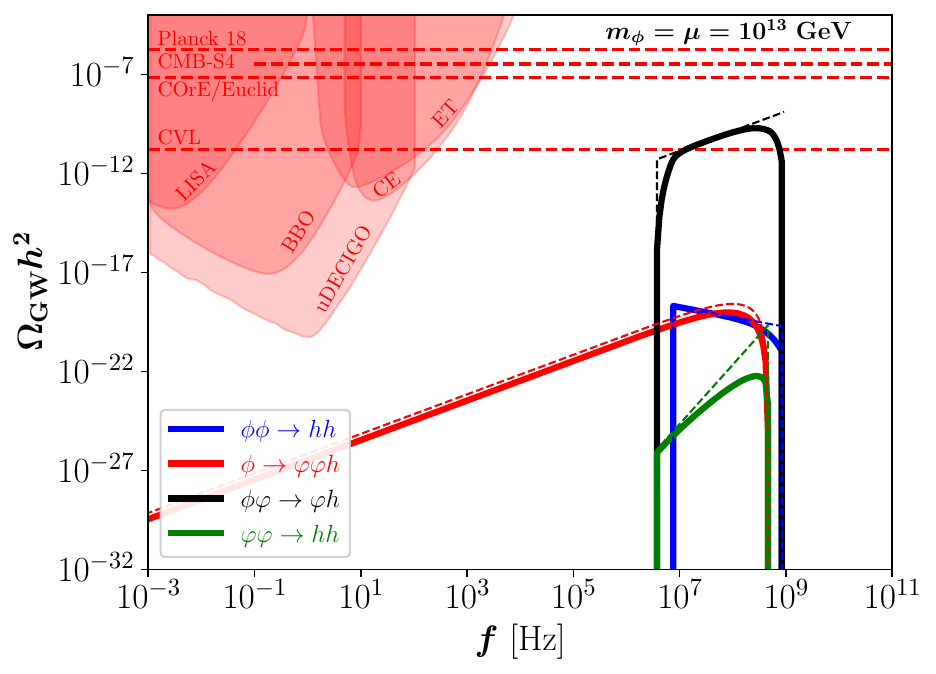}
		\caption{Comparison of the fully numerical spectra (thick solid lines) and the analytical approximations (thin dashed lines), for $m_\phi = \mu = 10^{13}$~GeV and $H_I = 2.0 \times 10^{-5}~M_P$.}
		\label{fig:compare2}
	\end{figure} 
	Figure~\ref{fig:compare2} presents a comparison between the fully numerical computation of GW spectra (thick solid lines) and the corresponding approximate analytical expressions (thin dashed lines) for $m_\phi = \mu = 10^{13}$~GeV and $H_I = 2.0 \times 10^{-5}~M_P$. Very good agreement is found.
	
	\bibliographystyle{JHEP}
	\bibliography{biblio}
\end{document}